\documentclass[fleqn,usenatbib]{mnras}

\usepackage{newtxtext,newtxmath}
\usepackage[T1]{fontenc}
\usepackage{ae,aecompl}

\usepackage{graphicx}	% Including figure files
\usepackage{amsmath}	% Advanced maths commands
\usepackage{amssymb}	% Extra maths symbols
\usepackage{array}
\usepackage{multirow}
\usepackage{tabularx}

\usepackage[dvipsnames]{xcolor}
\colorlet{orange}{green!10!orange!90!}

\usepackage[normalem]{ulem} %for strikeout /sout

\newcommand{\sunpc}{\,\text{M}_{\odot} \, \text{pc}^{-2}}	% solar masses per parsec-squared
\title[Molecular clouds in M51]{Molecular clouds in M51 from high-resolution extinction mapping}

\author[H. Faustino Vieira et al.]{Helena Faustino Vieira$^{1}$\thanks{E-mail: faustinovieirah@cardiff.ac.uk},
Ana Duarte-Cabral$^{1}$,
Timothy A. Davis$^{1}$,
Nicolas Peretto$^{1}$,
Matthew W. L. Smith$^{1}$,
\newauthor
Miguel Querejeta$^{2}$,
Dario Colombo$^{3}$,
Michael Anderson$^{1}$.
\\
$^{1}$ Cardiff Hub for Astrophysics Research and Technology (CHART), School of Physics \& Astronomy, Cardiff University, The Parade, CF24 3AA Cardiff, UK\\
$^{2}$ Observatorio Astronómico Nacional (IGN), C/ Alfonso XII 3, E28014 Madrid, Spain \\
$^{3}$ Max Planck Institute for Radioastronomy, Auf dem Hügel 69, D-53121 Bonn, Germany \\
}

\date{Accepted XXX. Received YYY; in original form ZZZ}

\pubyear{2023}

\begin{document}
\label{firstpage}
\pagerange{\pageref{firstpage}--\pageref{lastpage}}
\maketitle

\begin{abstract}

Here we present the cloud population extracted from M51, following the application of our new high-resolution dust extinction technique to the galaxy (Faustino Vieira et al. 2023). With this technique, we are able to image the gas content of the entire disc of M51 down to 5 pc ($0.14"$), which allows us to perform a statistical characterisation of well-resolved molecular cloud properties across different large-scale dynamical environments and with galactocentric distance. We find that cloud growth is promoted in regions in the galaxy where shear is minimised; i.e. clouds can grow into higher masses (and surface densities) inside the spiral arms and molecular ring. We do not detect any enhancement of high-mass star formation towards regions favourable to cloud growth, indicating that massive and/or dense clouds are not the sole ingredient for high-mass star formation. We find that in the spiral arms there is a significant decline of cloud surface densities with increasing galactocentric radius, whilst in the inter-arm regions they remain relatively constant. We also find that the surface density distribution for spiral arm clouds has two distinct behaviours in the inner and outer galaxy, with average cloud surface densities at larger galactocentric radii becoming similar to inter-arm clouds. We propose that the tidal interaction between M51 and its companion (NGC 5195) - which heavily affects the nature of the spiral structure - might be the main factor behind this. 

%s

\end{abstract}


%%%%%%%%%%%%%%%%%%%%%%%%%%%%%%%%%%%%%%%%%%%%%%%%%%

%%%%%%%%%%%%%%%%% BODY OF PAPER %%%%%%%%%%%%%%%%%%

\section{Introduction}

Stars form in the cold and dense molecular phase of the interstellar medium (ISM) in galaxies. The mechanism (or mechanisms) that trigger and regulate star formation (SF) in galaxies is still not well understood. In particular, it is not clear if the galactic environment has a direct impact in the galaxy's ability to form stars. Locations in galaxies with a higher density of molecular gas (e.g. spiral arms) seem to also harbour a higher concentration of young stars, which implies a higher star formation rate (SFR) towards those regions \citep[e.g.][]{bigiel_star_2008,schinnerer_pdbi_2013,leroy_molecular_2013}. One possible explanation for the higher SFR seen towards spiral arms is that the spiral arms themselves enhance SF. In a scenario first proposed by \cite{fujimoto_motion_1968} and \cite{roberts_large-scale_1969}, SF is triggered as the gas is compressed due to a shock that forms along the trailing edge of a spiral arm. Naturally, in this scenario, the "star formation efficiency" (SFE), or the SFR per unit gas mass, is higher in spiral arms than in less dense regions of galaxies \citep[e.g.][]{lord_efficient_1990,seigar_test_2002,silva-villa_relation_2012,yu_spiral_2021}. On the other hand, the increase of SFR towards spiral arms may just be a byproduct of the higher surface densities observed in that particular galactic environment. In other words, the underlying gravitational potential of the spiral reorganises and gathers the gas together, with no direct effect in the process of SF \citep[e.g.][]{elmegreen_density_1986}. If so, the observed SFE across galaxies should be effectively constant, which is in fact observed by several studies \citep[e.g.][]{leroy_star_2008, foyle_arm_2010, moore_effect_2012, ragan_prevalence_2016,urquhart_sedigism-atlasgal_2020, querejeta_stellar_2021}. Additionally, even if SF is not directly enhanced by the galactic environment, the large-scale dynamics may still play a critical role in regulating and disrupting SF across galaxies. Whether this dominates over other disruption mechanisms such as stellar feedback (and where this occurs in the galactic context) is still an active area of research \citep[e.g.][]{meidt_short_2015,chevance_lifecycle_2020,liu_wisdom_2021,chevance_pre-supernova_2022,liu_wisdom_2022,choi_wisdom_2023}.

Oftentimes, the molecular ISM of a galaxy is divided by astronomers into discrete structures known as molecular clouds (MCs), in order to better understand the initial conditions of SF. It is possible to investigate the link (or lack of) between the small, cloud-scale physics and the overarching galactic dynamics by analysing any systematic differences between MCs situated within different large-scale dynamical structures in galaxies (i.e. spiral arms, inter-arm regions, bars, etc.). In other words, by comparing the different cloud populations within galaxies, we can begin to understand if a galaxy's morphology has a direct impact in its ability to form stars. There have been many statistical characterisations of MCs in the Milky Way \citep[MW; e.g.][]{solomon_mass_1987, elia_first_2013, urquhart_sedigism-atlasgal_2020,duarte-cabral_sedigism_2021,colombo_sedigism_2022}, which benefit from the relatively small distances involved and thus achieve higher spatial resolution. Still, when it comes to linking MC properties and large-scale dynamics, Galactic studies are intrinsically limited given the difficulty of pinpointing locations of clouds within the context of the Galaxy \citep[e.g.][]{colombo_sedigism_2022}. Molecular gas observations in other galaxies do not suffer from this issue but instead are limited by sensitivity and resolution. However, with the advancement of instrumentation, extragalactic SF studies are now able to distinguish and resolve giant molecular clouds (GMCs), catapulting us into an exciting era of SF and ISM studies \citep[e.g.][]{koda_dynamically_2009,hughes_comparative_2013, colombo_pdbi_2014, sun_cloud-scale_2018, sun__dynamical_2020, querejeta_dense_2019, rosolowsky_giant_2021}.

In \citealt{faustino_vieira_high-resolution_2023} (hereafter Paper I), we presented a new high-resolution dust extinction technique that utilises archival optical \textit{Hubble Space Telescope} (HST) data to retrieve parsec-scale dust (and gas) surface density maps for entire nearby galaxies. In Paper I, we applied this technique to M51 as our test-case (briefly described in §\ref{sec:data}). M51 (NGC 5194) is an excellent candidate for cloud studies, as it is nearby and face-on, with bright spiral arms. It is a galaxy with a vast amount of multi-wavelength ancillary data and observational studies \citep[e.g.][]{vigne_hubble_2006,meidt_radial_2008,koda_dynamically_2009,schinnerer_multi-transition_2010,mentuch_cooper_spatially_2012,schinnerer_pdbi_2013,miyamoto_influence_2014,querejeta_gravitational_2016,messa_young_2018,querejeta_dense_2019}, as well as numerical simulations studying its evolution and dynamics \citep[e.g.][]{toomre_galactic_1972,salo_n-body_2000,dobbs_simulations_2010,tres_simulations_2021}. Here, our high-resolution gas surface density map of M51 from Paper I is used to extract an extensive cloud catalogue (§\ref{sec:cloudpopulation}). We analyse the properties of our molecular cloud sub-sample across large-scale galactic environment (§\ref{sec:env}), as well as with galactocentric distance (§\ref{sec:radial}). We provide a summary of our findings in §\ref{sec:sum_conc}.

\section{Data}
\label{sec:data} % used for referring to this section from elsewhere

In Paper I, we presented a novel technique which retrieves measurements of dust extinction along each line-of-sight for entire disc galaxies at parsec-scales, using archival HST optical data (F555W or V-band). A detailed description of the technique can be found in the original paper, but we present a brief overview of how the technique works here. Our high-resolution dust extinction technique is adapted from Galactic extinction studies conducted in the infrared (IR) \citep[e.g.][]{bacmann_isocam_2000,peretto_initial_2009}, which measure dust attenuation against a reconstructed, smoothly varying stellar light map, rather than determine the extinction from individual stars of similar spectral type. We construct this stellar distribution map by applying a sizeable median filter ($\sim600$\,pc) to the HST V-band image, after the removal of bright point-like sources. Fundamentally, this extinction technique compares the observed V-band intensity of each pixel in the map against the intensity from the reconstructed stellar distribution, which mimics the total stellar light if there were no extinction. The attenuation caused by dust is measured through:

\begin{equation}
    \label{eqn:method}
    \tau_\text{V} = - \text{ln} \left( \frac{I_\text{V} - I_{\text{fg}}}{I_{\text{bg}}} \right),
\end{equation}

\noindent where $\tau_\text{V}$ is the optical depth of the HST V-band, $I_\text{V}$ is the observed V-band intensity, and $I_\text{fg}$ and $I_\text{bg}$ are the foreground and background fractions, respectively, of the reconstructed stellar light model relative to the absorbing medium (i.e. dust). We assume that the attenuating dust sits in a layer near the galaxy's mid-plane in a "sandwich"-like geometry, and that the dust follows the radial profile of the stellar light. Our technique includes a calibration for the dust/stars geometry assumption, through comparison with \textit{Herschel Space Observatory} \citep{pilbratt_herschel_2010} lower-resolution observations of dust emission \citep{davies_dustpedia_2017}, so that our extinction dust mass estimates (at the $36"$ \textit{Herschel} resolution) are consistent with those derived from dust emission.

The measured V-band optical depth can be converted to dust surface densities ($\Sigma_\text{dust}$ in $\sunpc$ units) through a dust mass absorption coefficient for the V-band ($\kappa_\text{V}$): ${\Sigma_\text{dust}=\tau_\text{V}/\kappa_\text{V}}$. In Paper I and in this work, we adopt {$\kappa_V=1.786$\,pc$^2$\,M$_\odot^{-1}$} from \cite{draine_interstellar_2003}. Additionally, we assume a dust-to-gas mass ratio of 0.01 to derive gas surface densities from the dust map. The reader is referred to Paper I for further details.

Following our application of this extinction technique to M51, we obtained a gas surface density map of the galaxy at a spatial resolution of $\sim5$\,pc ($0.14"$), with which we are able to study spatially resolved cloud populations across the galaxy. This statistical analysis of molecular clouds (MCs) in the different dynamical environments of M51 (and with galactocentric radius) is the focus of the present paper. We adopt an inclination of $22^\circ$ \citep[][]{tully_kinematics_1974,colombo_pdbi_2014_moment}, a position angle of $173^\circ$ \citep{colombo_pdbi_2014_moment}, and a distance of 7.6\,Mpc \citep{ciardullo_planetary_2002} for M51.

\section{Cloud population from high-resolution extinction method}
\label{sec:cloudpopulation}

The gaseous content of galaxies is a multiphase continuum, and therefore "clouds" are not naturally occurring structures. Still, dividing the ISM into discrete clouds is a well-established technique that allows us to analyse the conditions of a galaxy's ISM in a statistical manner. In order to study how the properties of M51's ISM vary as a function of the galaxy's large-scale dynamics and galactocentric distance, we must first extract clouds from our extinction-derived surface density map ($\Sigma$).

\subsection{\texttt{SCIMES} cloud decomposition}
\label{sec:extraction}

\begin{figure*}
    \centering
    \includegraphics[width=0.95\textwidth]{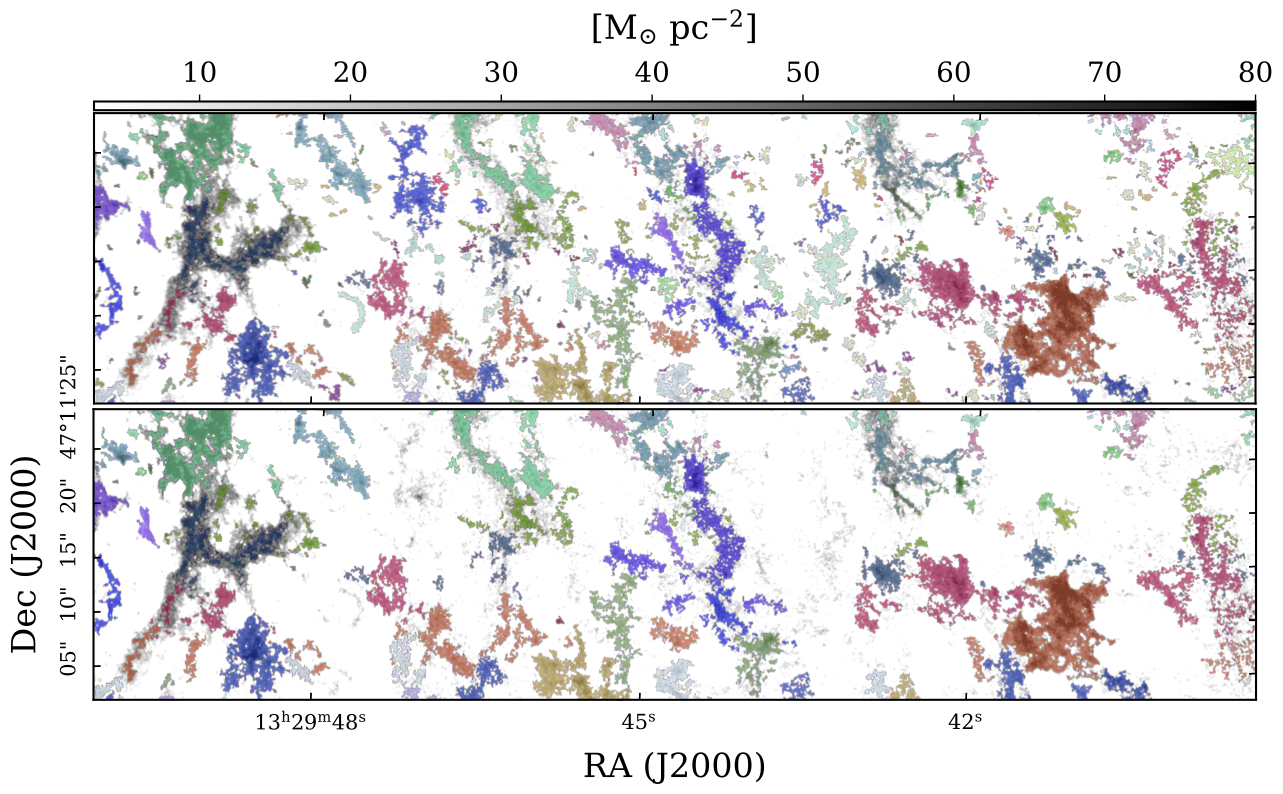}
    \caption{Example of the cloud extraction performed with \texttt{SCIMES}, where the cloud masks are overlaid with different, random colours in transparency, and the gas surface density is in the background greyscale. For the same region, the top panel shows the full \texttt{SCIMES} extraction (i.e. full cluster sample), and the bottom panel shows the sub-sample of molecular clouds as per our science cut ($\Sigma\geq10\sunpc$, $A\gtrsim90\,\text{pc}^2$ and $I_0\geq0.09\,\text{e}^-$/s, see text for details).}
    \label{fig:clouds}
\end{figure*}

The gas surface densities derived from the technique outlined in Paper I are decomposed into discrete clouds using the \texttt{SCIMES} clustering algorithm (v.0.3.2)\footnote{\url{https://github.com/Astroua/SCIMES/}}, initially described in \cite{colombo_graph-based_2015}. The updated version of \texttt{SCIMES} we use here is detailed in \cite{colombo_integrated_2019}. \texttt{SCIMES} works on the dendrogram tree of the input image - building the dendrogram from our gas surface density map is therefore the first step in our cloud extraction process.

A dendrogram \citep{rosolowsky_structural_2008} is comprised of three types of hierarchical structures: \textit{trunks} or "ancestors", which are the lowermost structures in the hierarchy of the input map from which all other structures in the dendrogram stem from; \textit{branches} which are the intermediate structures within the tree (i.e. structures that both have a parent and at least one child structure associated to them); and \textit{leaves}, the structures at the very top of the hierarchy with no child structures associated to them. In this study, we make use of the \texttt{ASTRODENDRO}\footnote{\url{http://www.dendrograms.org/}} implementation package to compute our dendrogram. \texttt{ASTRODENDRO} requires three initial inputs: \texttt{min\_value} (the minimum threshold below which no value is considered when building the dendrogram), \texttt{min\_delta} (the minimum difference between two peaks for the dendrogram to consider them as two separate, independent structures), and \texttt{min\_npix} (minimum number of pixels a structure must have to be considered an independent structure). To obtain the full dendrogram of our surface density map, we choose {\texttt{min\_value}\,$= 2\,\sunpc$}, {\texttt{min\_delta}\,$=9\,\sunpc$}, and {\texttt{min\_npix}\,$=27$\,pix} as our parameters. We choose a \texttt{min\_value} slightly above 0 to help segment the most diffuse material into trunks of a manageable size (if taking {\texttt{min\_value}\,$= 0\,\sunpc$}, the larger trunks would span almost the entire map). Tests were conducted in several small regions of M51 to check the effect of this lower threshold - no significant differences were observed on the final \texttt{SCIMES} extraction (except on the exact position of the boundaries of the most diffuse clouds), since most clouds are segmented above this threshold. Our choice of minimum value dismisses only 4.4\% of the total number of pixels in our map, which hold only 0.1\% of the total mass. To ensure that all structures within our dendrogram are spatially resolved, we set \texttt{min\_npix} to be roughly equal to the number of pixels equivalent to 3 resolution elements ($\sim9$ pixels per resolution element). We tested different values of \texttt{min\_delta}, from $3 \times$ the \texttt{min\_value} (i.e. $6\,\sunpc$) to $6\times$ the \texttt{min\_value} (i.e. $12\,\sunpc$), in various small regions of our map and found no significant difference in the final selection of structures, suggesting that the \texttt{SCIMES} segmentation outputs are not strongly impacted by the choice of dendrogram input parameters \citep{colombo_graph-based_2015}.

Using the dendrogram as a guide, \texttt{SCIMES} uses graph theory to perform spectral clustering and find regions with similar properties in emission (or in our case, in surface density) \citep{colombo_graph-based_2015}. In practice, \texttt{SCIMES} creates a graph that connects all leaves of the dendrogram (even those that do not have the same parent trunk) to build an affinity matrix that quantifies the relationship strength between the leaves. This process becomes extremely computationally (and memory) intensive when applied to large maps such as ours. To make cloud extraction more manageable, splitting the map into smaller sections is necessary. The most common way of doing this is to apply straight cuts to the data, which then requires dealing with clouds that touch those sharp edges separately \citep[as e.g.][]{colombo_integrated_2019, duarte-cabral_sedigism_2021}. We adopt a different approach and define "organic" masks using the trunk structures from the dendrogram, since these structures are at the bottom of the hierarchy and encompass all other structures present in the data. From the full dendrogram, we retrieve 29752 trunks in total. Ancestors that have just one child structure and ancestors that have no children structures (i.e. isolated leaves) cannot be clustered and therefore bypass the need to run the \texttt{SCIMES} clustering algorithm on them - they can directly be considered clouds. We then retrieve the masks of the remaining (3406) ancestor structures, and sort them into 4 horizontal strips of $0.03 ^{\circ}$ ($\sim 2$ arcmin), according to the Declination of their centroid position. This also allows us to create 4 non-overlapping sub-fields of our gas surface density map that, alongside the dendrogram, can be fed to \texttt{SCIMES}. For the cloud extraction with \texttt{SCIMES}, we opt to use the "radius" criterion for the clustering, with a user-defined scaling parameter of 90 pc (about two times the typical MW GMC size, e.g. \citealt{blitz_giant_1993}), to aid \texttt{SCIMES} on the identification of structures of a few tens of parsecs equally across the 4 fields and make the cloud extraction more robust\footnote{If left to decide the scaling parameter on its own, \texttt{SCIMES} works out the number of clusters to assign based on the contrast of the affinity matrices by default. As such, any given structure can change the way the dendrogram leaves are clustered depending on the dynamic range of the dataset. The dynamic range present within structures in the complex inner parts of M51 will be very different from the range present in the more diffuse outer parts. Therefore, in regions that span more hierarchical levels, the \texttt{SCIMES} extraction could potentially differ from the more "flat" regions (i.e. outskirts) without defining a common scaling parameter, making the clustering non-comparable between regions.}  The \texttt{SCIMES} segmentation recovered a total of 25291 clusters across the 4 sub-fields. Including the smaller ancestors that were directly put aside from the original dendrogram, our full sample has 51633 clouds. We produce a catalogue with the properties of our full cloud sample as well as the cloud assignment map for M51 which are made available at \url{https://dx.doi.org/10.11570/23.0030}. The cloud properties held in our catalogue are detailed in Appendix \ref{sec:appA}.

\subsection{Sub-sample of molecular clouds}
\label{sec:science_cut}

Stars are known to form in the coldest and densest (i.e. molecular) phase of the ISM. To establish any link between SF and galactic dynamics it is, therefore, necessary to focus on the structures encompassed in the star-forming molecular gas. Our cloud catalogue (§\ref{sec:extraction}) makes no distinction between atomic and molecular clouds since dust traces the total gas and we did not impose any restrictions on the cloud extraction itself. To retrieve a molecular sub-sample we must impose a surface density threshold from which we expect the ISM to be dominantly molecular. Consequently, we consider only the clouds with average surface density above $10 \, \sunpc$ \citep[e.g.][]{bigiel_star_2008, saintonge_cold_2022} as molecular clouds. To make sure the molecular clouds are well-resolved, we also impose that its footprint area be larger than 3 beams ($\sim$27.75\,pix, or an area $A$ of roughly $90\,\text{pc}^2$). Finally, our technique works out the dust attenuation through comparison with a reconstructed, smoothed stellar background. Therefore, structures that are picked up in regions with a faint background are not likely to be as well-defined as clouds in areas where the stellar background is more robust. We adopt a robust background threshold of $I_0=0.09$\,e$^{-}$/s (justification of this choice in Appendix \ref{sec:appA}). Each of these criteria has a corresponding flag in our full cluster catalogue: \textit{Molecular\_cut}, \textit{Size\_cut}, and \textit{Robust\_bg} (see Table \ref{tab:catalogue}). The resulting sub-sample of molecular clouds (which we will refer to as {\it{science sample}} from here on) contains 13258 molecular clouds, which are flagged in our full cluster catalogue with \textit{Molecular\_cut}=1, \textit{Size\_cut}=1, and \textit{Robust\_bg}=1. The bottom panel of Fig.~\ref{fig:clouds} shows the molecular clouds retrieved for a small section of M51 versus the full sample of clouds for the same region (top panel).

The total mass in our extinction-derived gas map of M51 is ${\text{M}_\text{gas}=8.9\times10^8\,(\pm3.4\times10^5)\,\text{M}_\odot}$\footnote{The calculated uncertainty on our total gas mass is derived from propagating the uncertainties of the opacity (and consequently mass) estimates for each pixel in our map (see Paper I). This, of course, is only the "formal" error, and is likely a lower estimate as it does not account for uncertainties in the assumed distance to the galaxy, opacity law, or any other systematic errors and assumptions.}. If we consider only the predominantly molecular gas in our map of M51 (i.e. $\Sigma > 10\,\sunpc$), we obtain a total molecular mass of ${\text{M}_{\text{mol}}=6.9\times10^8\,(\pm8.9\times10^4)\,\text{M}_\odot}$. Our full sample of clouds encompasses $\sim80\%$ of the total gas in our map of M51, whilst our science sample holds $\sim64\%$.

\begin{figure}
    \centering
    \includegraphics[width=0.4\textwidth]{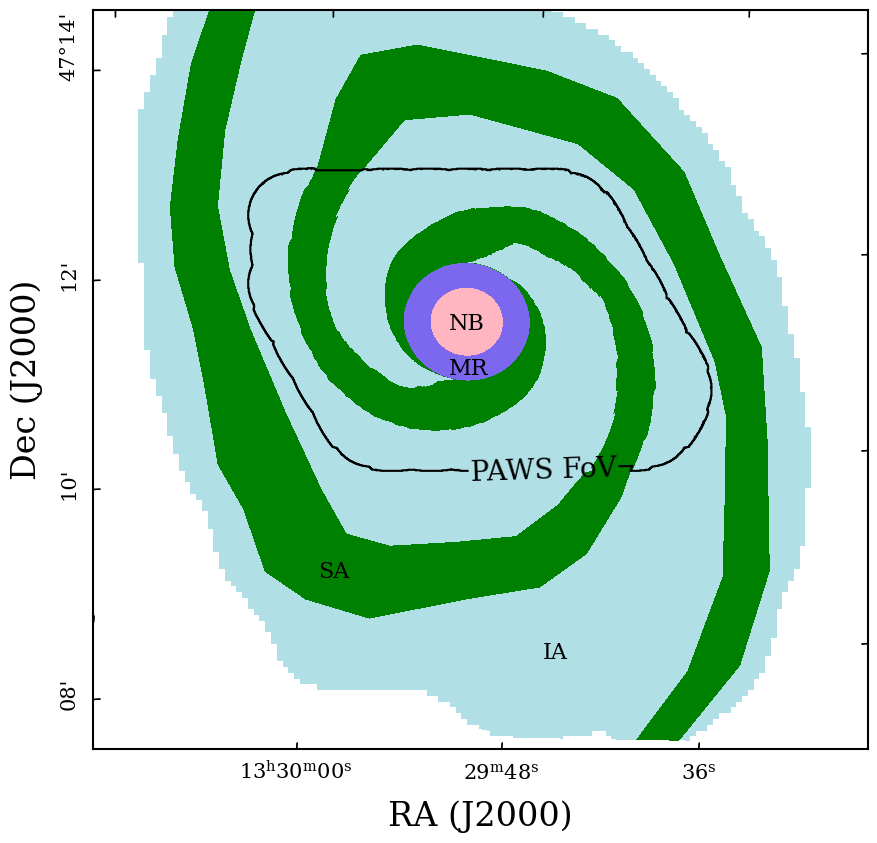}
    \caption{Environmental mask of M51 with the nuclear bar (NB) in pink, molecular ring (MR) in purple, spiral arms (SA) in green, and inter-arm regions (IA) in light blue. The PAWS FoV (field-of-view) is depicted as a black line contour.}
    \label{fig:env_mask}
\end{figure}

\section{Trends with large-scale environment}
\label{sec:env}

It is unclear if SF is more efficient in particular regions of galaxies, such as well-defined spiral arms, or if the higher rates of SF seen towards certain regions are simply a natural consequence of material crowding in the arms. If SF is dependent on the environment, then we would expect the cloud populations of those regions to have systematic differences in their characteristics. Some studies report some dissimilarities between their spiral arms and inter-arm populations \citep[e.g.][]{koda_dynamically_2009, colombo_pdbi_2014,pettitt_how_2020}, whilst other detect no significant differences in the global properties of the cloud population \citep[e.g.][]{duarte-cabral_what_2016, tres_simulations_2021, querejeta_stellar_2021}. The high resolution of our extinction map of M51 ($\sim 5$ pc) provides us with a unique opportunity to perform an in-depth statistical characterisation of MCs across different dynamical environments.

In order to analyse the environmental dependency of MC properties across the entire disc of M51 we must first construct a mask with the different large-scale environments. We approximate the inter-bar and nuclear bar of M51 (NB) to a circle with galactocentric radius\footnote{The galactocentric radius, $R_{\text{gal}}$, is the deprojected distance to the galactic centre, accounting for the inclination and position angle of M51 (see App.\,\ref{sec:appA}).} $R_{\text{gal}} < 0.85$ kpc, and the molecular ring (MR) to a ring spanning $0.85 < R_{\text{gal}} < 1.3$ kpc \citep{colombo_pdbi_2014}. We make use of the M51 environmental masks \citep[][see their Fig. 2]{colombo_pdbi_2014} from the PdBI Arcsecond Whirlpool Survey \citep[PAWS;][]{pety_plateau_2013,schinnerer_pdbi_2013} of the inner 5 kpc of the spiral arms (SA) and inter-arms (IA), and expand them for the full disc. This is done by using the extinction surface densities (convolved with a $\sim16"$ median filter) as a guide to continue the spiral arms from the end of the PAWS mask until the edges of M51. %, in a very simple by-eye fashion. 
Given that these masks were done mostly manually, they should not be taken as a strict or accurate definition of the spiral arm positions, instead, they serve as a means to provide all-galaxy statistical estimates. The resulting M51 environmental masks are shown in Fig.~\ref{fig:env_mask}.

\begin{figure}
    \centering
    \includegraphics[width=0.4\textwidth]{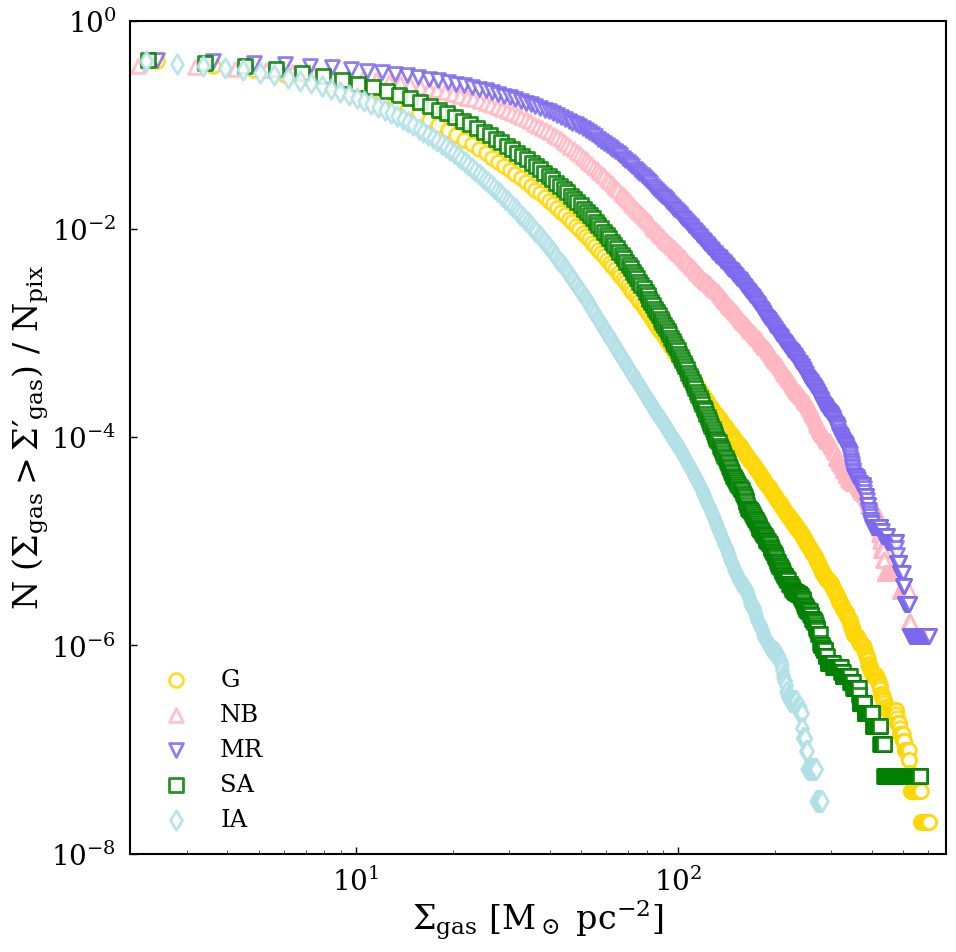}
    \caption{Cumulative distributions of the gas mass surface densities from the full map of M51, $\Sigma_{\text{gas}}$. The different colours represent the different environments: full galaxy (G) in yellow, nuclear bar (NB) in pink, molecular ring (MR) in purple, spiral arms (SA) in green, and inter-arms (IA) in blue. The y-axis, $N(\Sigma_\text{gas}>\Sigma'_\text{gas})$, is the fraction of pixels with surface density greater than a given value. All distributions are normalised by the number of pixels of the relevant environment, $N_{\text{pix}}$. Saturated pixels, i.e. pixels whose column we are not able to retrieve since their observed intensity is lower than the assumed foreground emission, are not considered in this plot.}
    \label{fig:sd_env}
\end{figure}

\subsection{Surface density probability density functions}
\label{sec:pdf}

Using the CO(2-1) observations from the PHANGS-ALMA survey \citep[][]{leroy_phangs-alma_2021}, \cite{sun_molecular_2020} and \cite{querejeta_stellar_2021} both report higher gas surface densities towards the centre of galaxies, with a more pronounced increase for barred galaxies. This was attributed to gas inflows driven by bars. 

Using our higher-resolution surface density map of M51, we investigate these trends in M51. Figure \ref{fig:sd_env} showcases the reverse cumulative distribution, or probability density function (PDF), of the gas mass surface densities for each environment (normalised by the number of pixels in each environmental mask). It is clear from the figure that the centre of M51 (NB + MR) hosts overall higher surface densities than the other regions, with the molecular ring in particular being the densest environment of M51, consistent with the results reported by the PHANGS-ALMA survey (as well as PAWS). In fact, the median of the surface density distribution ($\Sigma_{\text{gas}}$, listed in Table \ref{tab:cloud_properties}) for the molecular ring is over twice as large as the spiral arms value, and over 3 times the IA value. The molecular ring is effectively a dynamical gas transport barrier where material can accumulate easily, and produce the high densities observed \citep[e.g.][]{querejeta_gravitational_2016}. When compared to the MR, the nuclear bar $\Sigma_{\text{gas}}$ distribution displays a lack of intermediate to high surface densities (80-150 $\sunpc$), hinting at some disruptive mechanism that is absent from the molecular ring (likely streaming motions/shear driven by the bar's potential). The $\Sigma_{\text{gas}}$ distribution in the inter-arms shows a steady decline past the $10 \, \sunpc$ molecular threshold, consistent with a diffuse region from which we would expect more atomic gas. In comparison, the spiral arms contain a much larger amount of low to intermediate surface densities (10-80 $\sunpc$), although with a steeper decline towards high surface densities.

\begin{table*}
    \centering
    \begin{tabular}{c c | c c |  c c  c  c  c  c  c c}
    \hline
    \hline
    % Name of columns
    Env. & Env. tag & $\Sigma_{\text{gas}}$ & $A$ & $N_{\text{MC}}$ & $n_{\text{MC}}$ & $R_{\text{eq}}$ & log M & $\Sigma_{\text{MC}}$ & $AR_{\text{a/b}}$ & $AR_{\text{MA}}$ & $L_{\text{MA}}$ \\
    % Units of columns
     & & $\sunpc$ &kpc$^{2}$ & & kpc$^{-2}$ & pc & M$_\odot$ & $\sunpc$ & & & pc \\
     % Number of columns
    (1) & (2) & (3) & (4) & (5) & (6) & (7) & (8) & (9) & (10) & (11) & (12) \\
     \hline
     \hline
     % Global
     Global & G & 8.62$_{3.95}^{16.0}$ & 163.5 & 13258 & 81 & 8.82$_{6.76}^{13.7}$ & 3.60$_{3.34}^{4.00}$ & 14.5$_{11.7}^{19.7}$ & 1.85$_{1.49}^{2.35}$ & 3.16$_{2.34}^{4.77 }$ & 28.9$_{19.9}^{50.6}$ \\
     % Nuclear bar
     Nuclear bar & NB & 20.2$_{8.86}^{36.9}$ & 1.95 & 207 & 106 & 10.9$_{7.72}^{21.8}$ & 3.93$_{3.57}^{4.45}$ & 21.0$_{14.0}^{30.3}$ & 2.05$_{1.62}^{2.66}$ & 3.32$_{2.37}^{5.53}$ & 36.1$_{23.5}^{82.1}$ \\
     % Molecular ring
     Molecular ring & MR & 23.6$_{10.2}^{45.1}$ & 2.6 & 314 & 120 & 9.47$_{6.7}^{15.5}$ & 3.87$_{3.58}^{4.31}$ & 25.8$_{16.6}^{39.8}$ & 1.93$_{1.51}^{2.50}$ & 2.75$_{2.0}^{4.18}$ & 28.9$_{18.1}^{52.4}$ \\
     % Spiral arms
     Spiral arms & SA & 10.8$_{4.85}^{20.3}$ & 58 & 6042 & 104 & 8.82$_{6.76}^{14.0}$ & 3.64$_{3.37}^{4.05}$ & 15.6$_{12.2}^{22.1}$ & 1.84$_{1.50}^{2.34}$ & 3.16$_{2.3}^{4.7}$ & 28.9$_{19.9}^{50.6}$ \\
     % Inter-arms
     Inter-arms & IA & 7.53$_{3.51}^{13.4}$ & 101 & 6695 & 66 & 8.76$_{6.68}^{13.3}$ & 3.55$_{3.31}^{3.93}$ & 13.5$_{11.4}^{17.3}$ & 1.85$_{1.49}^{2.34}$ & 3.18$_{2.36}^{4.87}$ & 28.9$_{19.9}^{48.7}$ \\
     \hline
    \end{tabular}
    \caption{Properties of the different environments and of the MCs within each environment of M51. (1) Large-scale environment. (2) Environment abbreviations/tags. (3) Gas mass surface density of the environment, $\Sigma_{\text{gas}}$. (4) Area of environment, $A$. (5) Number of MCs per environment, $N_{\text{MC}}$. (6) Number density of MCs per environment, $n_{\text{MC}}$. (7) Equivalent radius, $R_{\text{eq}}$. (8) Mass, log $M$. (9) Average gas mass surface density of MCs within the environment, $\Sigma_{\text{MC}}$. (10) Ratio between major and minor axis of the MCs, $AR_{\text{a/b}}$. (11) Medial axis aspect ratio, $AR_{\text{MA}}$. (12) Length of the medial axis, $L_{\text{MA}}$. For columns (3) and (7) - (12), the median of the relevant distribution is written, with the lower quartile (Q25) as the subscript, and the upper quartile (Q75) as the superscript.}
    \label{tab:cloud_properties}
\end{table*}

\subsection{Molecular cloud properties}
\label{sec:mcproperties}

\begin{figure*}
    \centering
    \includegraphics[width=0.8\textwidth]{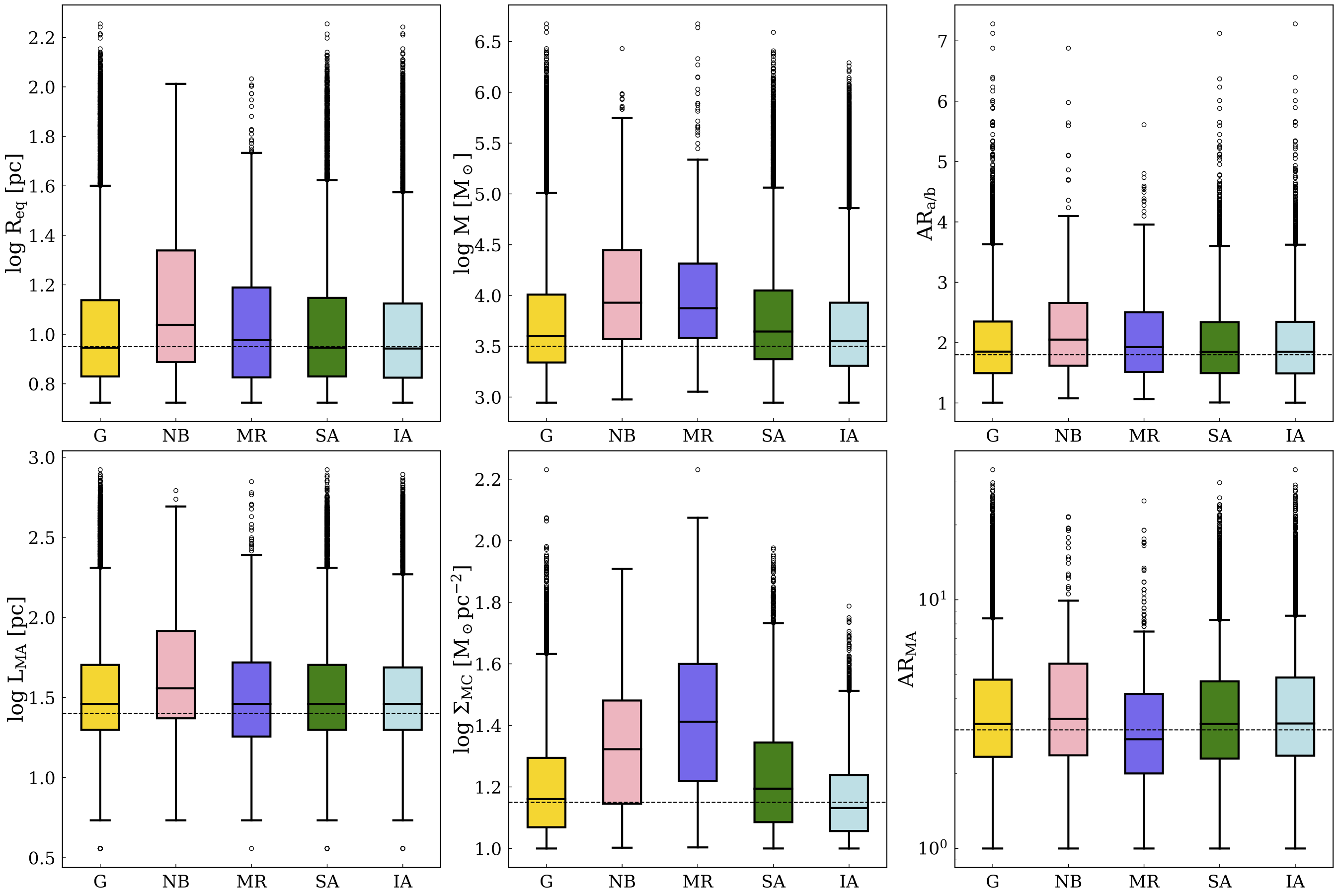}
    \caption{Boxplot representation of properties of the MCs in our science sample for the different large-scale environments of M51 (full galaxy in yellow, nuclear bar in pink, molecular ring in purple, spiral arms in green, and inter-arms in blue): equivalent radius $R_{\text{eq}}$ (\textit{top left}), mass $M$ (\textit{top middle}), major/minor axis ratio $AR_{\text{a/b}}$ (\textit{top right}), medial axis length $L_{\text{MA}}$ (\textit{bottom left}), average surface density of MCs $\Sigma_{\text{MC}}$ (\textit{bottom middle}), and medial axis aspect ratio $AR_{\text{MA}}$ (\textit{bottom right}). The lower and upper whiskers of the boxplot indicate respectively the minimum and maximum value of the dataset, and the coloured box illustrates the interquartile spread of the distribution (i.e. from 25th to 75th percentile), with the solid black line within the box being the median. Circles represent the outliers of the distributions. Dashed lines are reference lines set at arbitrary values for better visualisation.}
    \label{fig:boxplot}
\end{figure*}

Since MCs are not isolated and perfectly spheroidal structures, the complex dynamics of their surroundings will reflect on the shape and size of the clouds. If there are systematic variations of cloud morphology (as well as mass) between large-scale environments, this could shine a light on the dynamics at play and their impact on the formation and evolution of clouds. In their study of GMCs in M51, \cite{koda_dynamically_2009} propose an evolutionary picture: the spiral arm potential well encourages the molecular gas to consolidate into massive giant associations, which are then stretched apart and fragmented into smaller, lower-mass, elongated structures as they exit the spiral arms and encounter intense shear \citep[see also][]{vigne_hubble_2006}. This picture is supported by several other observational and numerical studies that report an abundance of filamentary objects in the inter-arms \citep[e.g.][]{ragan_giant_2014, duarte-cabral_what_2016, duarte-cabral_evolution_2017}, and high-mass objects in the spiral arms \citep[e.g.][]{dobbs_properties_2011,miyamoto_influence_2014, colombo_pdbi_2014}. It is important to note the effect of resolution in these findings however, since lower resolution can blend structures into massive associations, notably in crowded regions like the spiral arms. In another study of M51, \cite{meidt_short_2015} found that both shear driven by galactic dynamics and stellar feedback can be responsible for disrupting MCs, and consequently suppressing SF. More recently, \cite{chevance_pre-supernova_2022} argue that early (pre-supernovae) stellar feedback mechanisms are the main driver of cloud disruption in galaxies. Determining which is the dominant process in SF regulation (shear or stellar feedback), and where in galaxies this occurs, is crucial to better understand cloud lifecycles and lifetimes, and their role in SF and in galaxy evolution \citep[e.g.][]{kruijssen_fast_2019}. With our dust extinction technique, we are able to break up large cloud associations and resolve MC structure with a lot more detail, and thus observe the impact of these disruption mechanisms on individual clouds. In this work, we examine the properties of MCs in search of systematic differences between large-scale environments, which would suggest a direct link between cloud-scale physics and galactic environment. Presently, we do not attempt to pinpoint the exact driver of different characteristics in cloud populations (i.e. driven by shear or stellar feedback), but our spatially resolved cloud catalogue does allow for such an exercise. In future work, we plan to also analyse cloud properties with azimuth and as a function of distance to the nearest spiral arm, and also in relation to various SF tracers.

The various cloud properties analysed in this paper are listed in Table \ref{tab:cloud_properties} (and further detailed in Appendix \ref{sec:appA}) with some of them also illustrated in Fig.~\ref{fig:boxplot}. We find that the central region of M51 shows systematic differences from the characteristics of the MCs from the disc, with most of the analysed properties presenting higher median values in the centre. This suggests that M51's centre has a substantial impact in the formation and evolution of all of its MCs \citep[also noted by][]{querejeta_stellar_2021}. In particular, MCs located in the molecular ring tend to be denser, whilst in the nuclear bar they are more elongated (but equally massive). On the other hand, the spiral arm and inter-arms MC populations do not show significant differences in their statistics, with the exception of the average cloud surface densities and mass, where the SA median is slightly higher. In their simulation of an M51 analogue, \cite{tres_simulations_2021} find similar trends in cloud properties; i.e. the central clouds show significantly different characteristics from the disc, whilst the SA and IA cloud populations seem very similar in their properties.

\subsubsection{Mass and surface density}

As can be seen in Table \ref{tab:cloud_properties}, our molecular clouds have a median mass of roughly ${4\times10^3\,\text{M}_\odot}$ and a radius of about 9\,pc. These are smaller clouds (in both size and mass) than those from the numerical work of \cite{tres_simulations_2021}, with a median mass and radius of ${2\times10^4\,\text{M}_\odot}$ and 16\,pc, respectively. This could be due to the model's resolution limitations in the lower column density regime \citep[see Fig. 3 of ][]{tres_simulations_2020}, coupled with the specific prescription for the Supernovae feedback, which naturally leads to a lower amount of MCs with smaller masses and radii in low column regions such as the inter-arms. Nevertheless, we see similar trends in cloud properties between environments and the same range in cloud mass values (extending up to about $10^{6.5}\,\text{M}_\odot$) as \cite{tres_simulations_2021}. Our MCs are also much smaller on average than the PAWS clouds \citep{colombo_pdbi_2014}, where the median mass and radius are ${7.6\times10^5\,\text{M}_\odot}$ and 48\,pc, respectively. It is clear that comparing cloud masses between different studies is not the most informative, as masses (and sizes) are heavily dependent on each study's definition of a molecular cloud and its boundaries, as well as resolution limits which might lead to beam smearing in crowded regions, as well as undetections. In fact, given its $1"$ resolution ($\sim40$\,pc), the completeness limit of the PAWS catalogue, ${3.6\times10^5\,\text{M}_\odot}$, is already much higher than our median cloud mass. Similarly, the resolution limitations imply a minimum effective radius of 20\,pc for the PAWS GMCs, which is already over a factor 2 larger than our typical cloud radius ($\sim9$\,pc), meaning the average MC in our catalogue might go undetected in PAWS or appear unresolved within a beam area. Therefore, care is needed when comparing cloud catalogues, especially when comparing absolute values.

\begin{figure}
    \centering
    \includegraphics[width=0.45\textwidth]{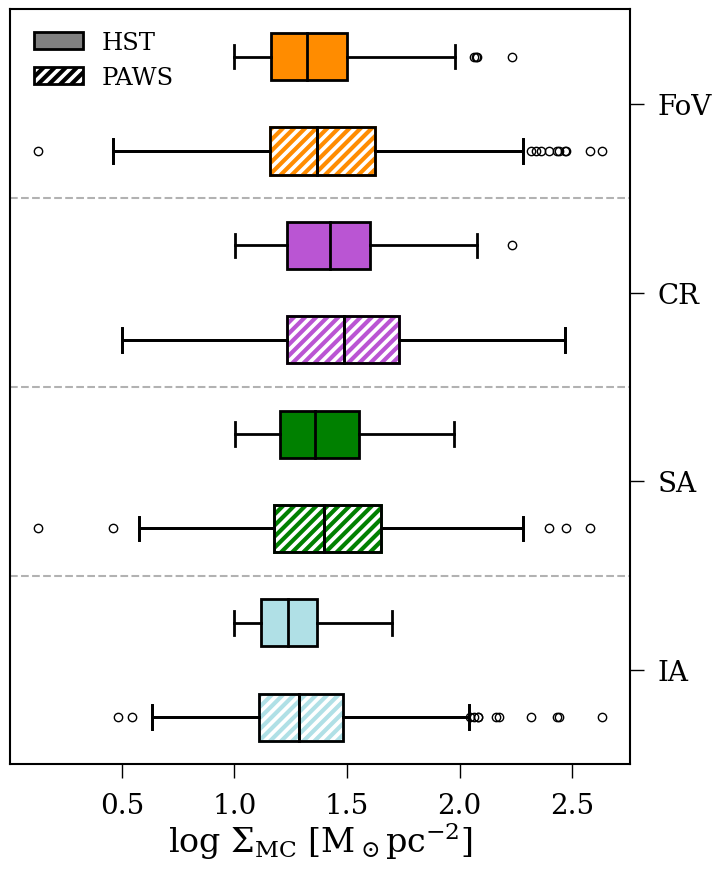}
    \caption{Boxplot representation of average cloud surface densities, $\Sigma_\text{MC}$, for the cross-matching subsets of our extinction-derived HST clouds (solid) and PAWS GMCs (hatched). The different large-scale environments are represented with the following colour-scheme from top to bottom: PAWS FoV in orange, central region (CR, encompasses nuclear bar and molecular ring) in violet, spiral arms in green, and inter-arms in blue. The lower and upper whiskers of the boxplot indicate respectively the minimum and maximum value of the dataset, and the coloured box illustrates the interquartile spread of the distribution (i.e. from 25th to 75th percentile), with the solid black line within the box being the median. Circles represent the outliers of the distributions.}
    \label{fig:paws_hst_match}
\end{figure}

In Fig.\,\ref{fig:paws_hst_match} we present a comparison of average cloud surface densities for the cross-matches between the PAWS GMCs and our extinction-derived HST MCs. By definition, average cloud surface densities already account for cloud size (i.e. $\Sigma_{\text{MC}}=\text{M}/A$), which reduces the effect of different resolutions between studies, although it is not entirely removed as we will address later. We perform this cross-matching to ensure that we are only comparing clouds that roughly exist in the same space, so that the comparison is as fair as possible. The footprint masks of the PAWS GMCs \citep{colombo_pdbi_2014} were deprojected into 2D masks and regridded to the HST native grid ($0.049"$/pix). We find that 1296 of the total 1507 PAWS GMCs have a spatial match with HST clouds, meaning our catalogue successfully matches with $86\%$ of the PAWS catalogue. The remaining 211 unmatched PAWS GMCs are likely associated with clusters, which prevent any measurement of extinction in the relevant region (as detailed in Paper I). Out of the 4843 HST MCs in the PAWS FoV (highlighted in Fig.\,\ref{fig:env_mask}), only $35\%$ (1700) match with at least one GMC in PAWS. This significant fraction of unmatched HST clouds is again a reflection of the differences in column sensitivity and resolution of the two catalogues: the unmatched HST clouds typically have lower average surface density ($\sim15.1\,\sunpc$) and thus are likely associated with CO-dark molecular gas, and are also too small ($\sim8.5$\,pc) to be resolved by PAWS. In addition to cross-matching, we also recalculate the average surface densities of PAWS clouds with a scaled CO-to-H$_2$ conversion factor. \cite{colombo_pdbi_2014} employ the standard Galactic CO-to-H$_2$ conversion factor, ${X_{\text{CO}}=2\times10^{20}\,\text{cm}^{-2}\text{(K km s}^{-1})^{-1}}$, when deriving their cloud masses (and surface densities) from CO luminosity. In Paper I, we found that the determination of $X_{\text{CO}}$ is heavily influenced by the assumed dust model, and that assuming the Galactic $X_{\text{CO}}$ overestimates the PAWS surface densities by roughly a factor 7 relative to our estimates. Adopting the scaled value of ${X_{\text{CO}}=3.1\,(\pm0.3)\,\times10^{19}\,\text{cm}^{-2}\text{(K km s}^{-1})^{-1}}$ removes this discrepancy and makes the two studies comparable (see Paper I for more details). Once all these steps are taken to ensure the comparison of cloud properties between our catalogue and the one from PAWS is as fair as possible, we find that the median cloud surface densities (as well as observed trends between environments) are virtually identical for both catalogues (shown in Fig.\,\ref{fig:paws_hst_match}). Still, from the figure we can see that the observed range of cloud surface density values for the PAWS GMCs are consistently larger than the HST range. Again, this is likely linked to resolution: in crowded areas, a larger beam might blend multiple clouds in the same line of sight resulting in larger surface densities, whilst small-sized and more isolated clouds might get smeared within the beam, resulting in a "dilution" of the observed flux in a larger area (i.e. lower surface density).

\subsubsection{Elongation}

In order to evaluate the elongation of clouds, we employ two different methods of measuring aspect ratio. The first is a moments-based aspect ratio, $AR_{\text{a/b}}$, which is the ratio between a cloud's surface density-weighted semi-major axis ($a$) and semi-minor axis ($b$). The second metric is purely geometrical, based on the medial axis of the cloud, which is the longest continuous line connecting the points within a cloud furthest away from its external edges. The medial axis aspect ratio, $AR_{\text{MA}}$, is defined as the ratio between the length of the medial axis ($\text{L}_\text{MA}$) and twice the average distance from the medial axis to the cloud's edge ($\text{W}_\text{MA}$). For further details please refer to Appendix \ref{sec:appA}. Both aspect ratio metrics suggest the same trends between the spiral arms and inter-arms in that, although the values are higher overall for $AR_{\text{MA}}$, both environments seem to have equally elongated clouds. It is possible that any disparities between these two populations are only seen in the most extreme clouds (i.e. tails of the distributions) rather than the bulk of the population, which will be further examined in §\ref{sec:extreme}. Both $AR_{\text{MA}}$ and $AR_{\text{a/b}}$ indicate that MCs in the nuclear bar are more elongated than anywhere else in the galaxy. Due to their non-axisymmetric nature, bars are known to drive gas inflows and produce intense shear \citep[e.g.][]{meidt_gas_2013, querejeta_gravitational_2016}, likely stretching clouds apart (i.e. providing higher aspect ratios). The picture is more unclear when considering the molecular ring population; the overall trend with the remaining large-scale environments is different for the two metrics (for $AR_{\text{a/b}}$, the MR clouds have the second highest median, but for $AR_{\text{MA}}$, the same clouds have the lowest median). The difference in trends between the two aspect ratio metrics employed here is not wholly unexpected. In a recent morphology study of the SEDIGISM clouds \citep{duarte-cabral_sedigism_2021}, \cite{neralwar_sedigism_2022} note that measures of aspect ratio vary quite significantly depending on the methodology adopted. Using an aspect ratio as a proxy for cloud elongation is not straightforward as it depends on the specific morphology of the clouds, and should therefore be used with care (for an example see Appendix \ref{sec:appB}). It is beyond the scope of this paper to address this discrepancy between aspect ratios, but a more detailed cloud morphology study using RJ-plots \citep[][based on the morphological classification technique J-plots developed by \citealt{jaffa_j_2018}]{clarke_rj-plots_2022} is within our future plans.

\begin{figure}
    \centering
    \includegraphics[width=0.4\textwidth]{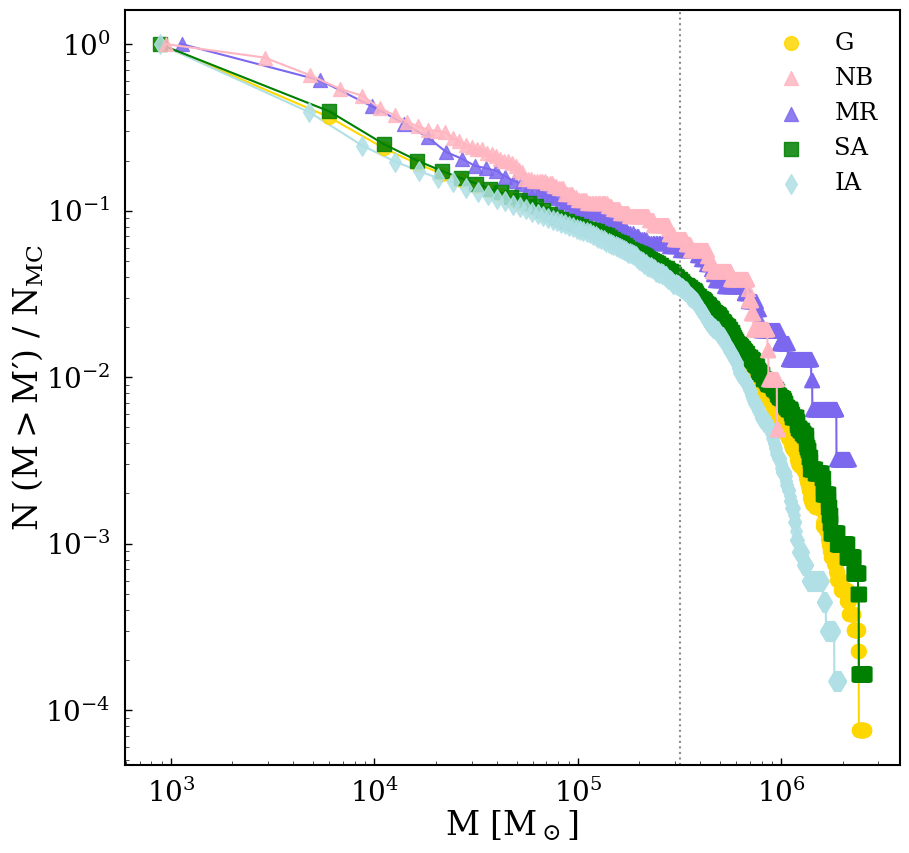}
    \caption{Cumulative mass distributions for the different environments in M51 (full galaxy in yellow, nuclear bar in pink, molecular ring in purple, spiral arms in green, and inter-arms in blue). The distributions are normalised by the number of clouds of each environment, $N_{\text{MC}}$ (listed in Table \ref{tab:cloud_properties}). Clouds with saturated pixels/uncertain opacities are not included.}
    \label{fig:mass_clouds}
\end{figure}

\subsection{Cloud cumulative mass distributions}
\label{sec:cmd}

\begin{table}
    \centering
    \begin{tabular}{c | c c c}
    \hline
    \hline
    \multicolumn{4}{c}{Full galaxy} \\
    \hline
    Env. & $\gamma$ & $M_0$ & $N_0$ \\
     & & $10^6$ M$_\odot$ & \\
    \hline
    G & -2.43 $\pm$ 0.01 & 2.21 $\pm$ 0.02 &  35.7 $\pm$ 1.04\\
    NB & -1.01 $\pm$ 0.02 & 1.20 $\pm$ 0.01 & ... \\
    MR & -1.28 $\pm$ 0.01 & 2.18 $\pm$ 0.02 & 24.5 $\pm$ 1.65 \\
    SA & -2.19 $\pm$ 0.01 & 2.42 $\pm$ 0.02 & 23.7 $\pm$ 0.57 \\
    IA & -2.61 $\pm$ 0.01 & 1.64 $\pm$ 0.01 & 18.99 $\pm$ 0.68 \\
    \hline
    \multicolumn{4}{c}{PAWS FoV} \\
    \hline
    G & -2.13 $\pm$ 0.01 & 2.16 $\pm$ 0.01 & 32.2 $\pm$ 0.86 \\
    SA & -1.77 $\pm$ 0.01 & 2.22 $\pm$ 0.02 & 25.4 $\pm$ 1 \\
    IA & -2.38 $\pm$ 0.01 & 1.81 $\pm$ 0.01 & 12.7 $\pm$ 0.38 \\
    \hline
    \end{tabular}
    \caption{Parameters from truncated power-law fits ($\gamma$, $M_0$ and $N_0$) for the full galaxy (top section) and for the PAWS FoV (bottom section). The errors quoted are the standard deviations of the fits. The $N_0$ estimate for the nuclear bar is omitted from this table as it does not have any physical meaning. The $\gamma$ estimate for this region is nearly -1, the point for which Eq. (\ref{eqn:t_power_law}) is no longer valid, and consequently determining $N_0$ becomes impossible.}
    \label{tab:cumulative}
\end{table}

\begin{figure*}
    \centering
    \includegraphics[width=\textwidth]{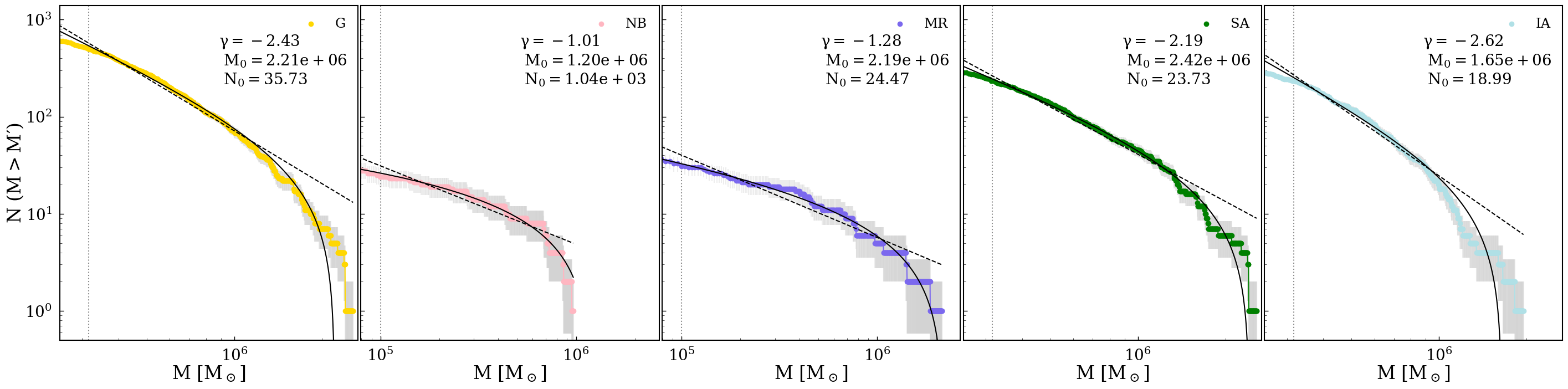}
    \caption{Cumulative mass distributions for the MCs in our science sample for the different large-scale environments (from left to right: global in yellow, nuclear bar in pink, molecular ring in purple, spiral arms in green, and inter-arms in blue), normalised by the area of each environment (listed in Table \ref{tab:cloud_properties}). Dotted grey lines depict the fiducial mass starting from which the fits were performed ($10^{5.5}$ M$_\odot$ for G, SA, IA, and $10^{5}$ M$_\odot$ for NB, MR). Dashed black lines represent the simple power-law fits performed, and solid black lines the truncated power-law fits. The environment label, as well as the spectral index $\gamma$, the maximum mass $M_0$ and $N_0$ for the truncated fits, are shown in the top right. The errorbars in light grey are the Poisson errors on the counts ($\sqrt{N}$). Clouds with saturated pixels are not included in these distributions.}
    \label{fig:mass_fits}
\end{figure*}

As highlighted in the previous section (§\ref{sec:mcproperties}), although there are clear differences in the masses of MCs between the centre and the disc of M51, the medians of the distributions alone are not the most informative, particularly when analysing any potential differences between the IA and SA clouds. Therefore, we also analyse how cloud masses are distributed within each large-scale environment, by building a cumulative mass distribution. To obtain the mass spectra for our sample, we opt to exclude the clouds that include saturated pixels, since these clouds have more uncertain masses (see Paper I). Figure\,\ref{fig:mass_clouds} shows the mass spectra for the MCs in our science sample for the different M51 environments normalised by the number of clouds in each environment. From the figure, it is possible to see that the central regions of M51 (NB and MR) have the highest concentration of high-mass clouds (M$\gtrsim 10^{5.5}$ M$_\odot$), followed by the spiral arms. There's also a sharp decline in high-mass objects for the IA - the MCs in this region seem to have predominantly low to intermediate masses. These findings agree with what was found, albeit at a lower resolution, by \cite{colombo_pdbi_2014} in their GMC study of M51 using PAWS CO data, and more recently by \cite{rosolowsky_giant_2021} in their study of GMCs across PHANGS spiral galaxies. The trends seen in the MC mass spectra follow the same trends as what we saw for the pixel-by-pixel surface density distributions (Fig.~\ref{fig:sd_env}), and thus the cloud segmentation process used in this study is unlikely to be the cause of the cloud mass distributions seen here. 

The cumulative mass distribution can be fit with a simple power-law of the shape: ${N = \left( M / M_0 \right)^{\gamma + 1}}$, where $N$ is the number of clouds with mass $M$ that is larger than the reference mass $M_0$, and $\gamma$ the index of the power-law. However, given the steepening of the mass distributions seen at higher masses, we opt for a truncated power-law of the form:

\begin{equation}
    N = N_0 \left[ \left( \frac{M}{M_0} \right)^{\gamma + 1} - 1 \right],
    \label{eqn:t_power_law}
\end{equation}

\noindent with $M_0$ being the maximum mass of the distribution, and $N_0$ the number of clouds corresponding to the truncation mass, ${M_t = 2^{1/(\gamma+1) M_0}}$ (i.e. the point at which the mass distribution stops following a simple power-law). The index of the truncated power-law informs us on how the mass is distributed: in massive cloud structures for $\gamma > -2$, and in smaller clouds for $\gamma < -2$. For the spiral arms and inter-arms, we fit the mass spectra with Eq. (\ref{eqn:t_power_law}) for masses greater than $10^{5.5}$ M$_\odot$, which is the point from which the distributions seem to have a shape similar to a truncated power-law. We adopt a lower mass threshold of $10^5$ M$_\odot$ for the nuclear bar and molecular ring due to the reduced number of clouds with masses higher than $10^{5.5}$ M$_\odot$. The resulting parameters from the fits are listed in Table \ref{tab:cumulative}, and the fits themselves (both simple and truncated) are shown in Fig.~\ref{fig:mass_fits}. 

The global cumulative mass distribution of all the MCs in our sub-sample is very steep, with a fitted index $\gamma < -2$, which indicates that our M51 clouds are preferentially low-mass objects. We can again see in Fig.~\ref{fig:mass_fits} that the cloud population in the centre of M51 (NB + MR) has different characteristics from the disc (SA + IA) with very different slopes of the truncated fits. Both the nuclear bar and molecular ring present $\gamma > -2$, whilst the spiral arms and inter-arms fits have $\gamma < -2$, suggesting that clouds in the disc are typically low-mass, whilst MCs in the centre have larger masses, in line with our results from §\ref{sec:mcproperties}.

Although the nuclear bar truncated fit has the shallowest slope (indicative of preference towards high-mass objects), the distribution itself does not extend to high masses (highest mass $\sim 9\times10^5$ M$_\odot$), suggesting that cloud growth is being hindered and/or that massive clouds are being destroyed in this region. This is likely a result of the complex dynamics and intense shear caused by the bar, although \cite{colombo_pdbi_2014} also argue that the enhanced interstellar radiation field in M51's bulge could also have an effect. On the other hand, the molecular ring also has a low spectral index but its distribution reaches higher mass values ($\sim 2\times10^{6}$ M$_\odot$), consistent with an environment that promotes cloud agglomeration.

The IA cumulative mass distribution presents the steepest slope out of all the considered environments, indicating that the inter-arms host dominantly lower mass MCs. Furthermore, the IA distribution extends up to a smaller mass relative to the spiral arms cumulative mass distribution, even though the two distributions are very similar in the low-to-intermediate mass range ($<10^{5.5}$ M$_\odot$, see Fig.~\ref{fig:mass_clouds}). It seems that high-mass objects in the inter-arms either have difficulty forming or are destroyed quickly after formation. On the other hand, the spiral arms mass cumulative distribution reaches the highest mass among all considered environments ($\sim 2.6\times10^6$ M$_\odot$), even though its slope is relatively steep. The SA then have favourable conditions for clouds to grow more massive even though most of its population seem to be low-mass objects. From their simulations of an interacting galaxy, \cite{pettitt_how_2020} also observe a steeper slope in their IA cumulative cloud mass distribution relative to the SA slope, with SA clouds reaching higher masses. Additionally, the fitted index of their whole cloud population, $\gamma = -2.39$, is very close to the value we find ($\gamma=-2.43$).

Interestingly, the fitted parameters ($\gamma$, $M_0$, and $N_0$) change quite significantly when fitting the mass cumulative distributions of only the MCs inside the PAWS FoV (i.e. clouds at smaller galactocentric radii, $R_{\text{gal}} \lesssim 5$ kpc), as can be seen from the bottom section of Table\,\ref{tab:cumulative}. Overall, MCs seem to be more massive inside the PAWS FoV than when considering the full galaxy, hinting at a radial trend in cloud mass (which will be analysed in more depth in §\ref{sec:radial}). Notably, the slope of the truncated fit for the inner spiral arms is much shallower than for the full arms, with $\gamma > -2$, whilst the index for the inner inter-arms is still $\gamma < -2$. 

\begin{figure*}
    \centering
    \includegraphics[width=0.9\textwidth]{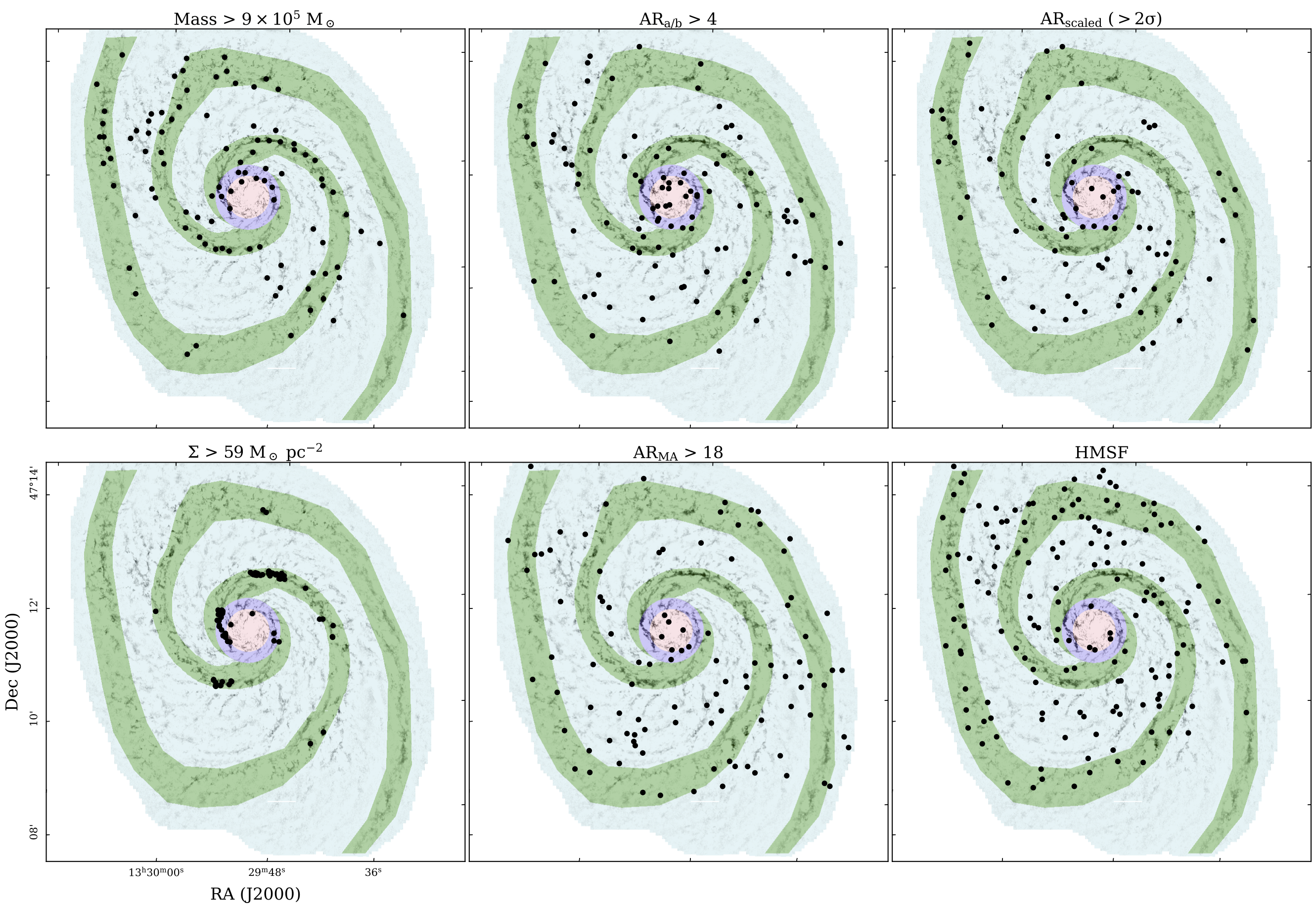}
    \caption{Distribution of the most extreme clouds in our science sample in terms of their mass $M$ (\textit{top left}), major/minor axis ratio $AR_{\text{a/b}}$ (\textit{top middle}), rescaled and standardised aspect ratio (above 2$\sigma$, \textit{top right}), average surface density $\Sigma_{\text{MC}}$ (\textit{bottom left}), medial axis aspect ratio $AR_{\text{MA}}$ (\textit{bottom middle}), and potential high-mass star formation (i.e. MCs with an associated $8\,\upmu$m source, \textit{bottom right}). Where relevant, at the top of each panel is the corresponding lower bound of the relevant property for the clouds shown.}
    \label{fig:extreme}
\end{figure*}

\subsection{Extreme clouds}
\label{sec:extreme}

\begin{table*}
    \centering
    \begin{tabular}{c|c|c|c|c|c|c|c|c|c}
    \hline
    \hline
    Env. &  $f$ & $f_{M}$ & $f_{\Sigma}$ & $f_{\text{AR}_\text{MA}}$ & $f_{\text{AR}_\text{a/b}}$ & $f_{\text{AR}_\text{scaled}}$ & $f_{\text{HMSF}}$ & $f_{\text{L}_\text{MA}}$ & $f_\text{A}$ \\
    (1) & (2) & (3) & (4) & (5) & (6) & (7) & (8) & (9) & (10) \\
    \hline
    NB & 0.02 & 0.03 & 0.02 & 0.05 & 0.1 & 0.03 & 0.02 & 0.02 & 0\\
    MR & 0.02 & 0.08 & 0.22 & 0.03 & 0.08 & 0.08 & 0.05 & 0.03 & 0.01 \\
    SA & 0.45 & 0.58 & 0.75 & 0.30 & 0.35 & 0.41 & 0.51 & 0.32 & 0.47 \\
    IA & 0.51 & 0.31 & 0.01 & 0.62 & 0.47 & 0.48 & 0.42 & 0.63 & 0.52 \\
    \hline
    \hline
    \multicolumn{2}{c |}{$\chi^2$} & 25.6 & 230 & 15.7 & 61.7 & 14.1$^{+}$ & 23.1$^{*}$ & 7.43 & 2.44 \\
    \multicolumn{2}{c |}{p$_{\text{rnd}}$} & 0.0004 & <10$^{-5}$ & 0.0036 & <10$^{-5}$ & 0.007$^{+}$ & 0.0001$^{*}$ & 0.06 & 0.48 \\
    \multicolumn{8}{l}{$^{+}$ estimated with $N = 88$ clouds.}\\
    \multicolumn{8}{l}{$^{*}$ estimated with $N = 460$ clouds.}\\
    \end{tabular}
    \caption{Extreme cloud (top 100) fractions across galactic environments. (1) Environment tag (\textit{Env}) for the nuclear bar (NB), the molecular ring (MR), the spiral arms (SA), and the inter-arms (IA). (2) Fraction of clouds of each environment ($N_\text{env}$) with respect to the total number of clouds in the science sample ($N_\text{MC}$), $f$ ($f=N_\text{env}/N_\text{MC}$). (3) Fraction of most massive clouds, $f_M$. (4) Fraction of clouds with the highest surface density, $f_\Sigma$. (5) Fraction of most elongated clouds according to the medial axis aspect ratio, $f_{\text{AR}_\text{MA}}$. (6) Fraction of most elongated clouds according to the moment aspect ratio, $f_{\text{AR}_\text{a/b}}$. (7) Fraction of most elongated clouds ($>2\sigma$) according to both metrics of aspect ratio scaled and standardised, $f_{\text{AR}_{\text{scaled}}}$. (8) Fraction of high-mass star forming clouds which have an associated 8\,$\upmu$m source (including both exact and closest match) from \protect\cite{elmegreen_highly_2019}, $f_{\text{HMSF}}$. (9) Fraction of longest clouds according to their medial axis length, $f_{\text{L}_{\text{MA}}}$. (10) Fraction of largest clouds in terms of their footprint area, $f_A$. The bottom portion of the table shows the results from our investigation into the significance of the statistical difference in the distribution of our extreme clouds compared to the global science sample. We list the $\chi^2$ value between each distribution of extreme clouds and the full science sample. $\text{p}_{\text{rnd}}$ represents the probability or likelihood of obtaining the listed $\chi^2$ values from a pure random draw of $N=100$ clouds ($N=88$ and $N=460$ for the $\text{AR}_\text{scaled}$ and HMSF sub-samples, respectively) from our science sample.}
    \label{tab:extreme}
\end{table*}

As evidenced by §\ref{sec:mcproperties} and §\ref{sec:cmd}, although the bulk properties of a galaxy's different cloud populations may be fairly similar, differences arise when analysing the tails of the distributions \citep[see also][]{duarte-cabral_what_2016,duarte-cabral_sedigism_2021}. If "extreme" clouds (i.e. the clouds at the tail of the relevant distribution) are enhanced in certain large-scale galactic environments, then this points at physical processes that directly facilitate the formation of specific types of clouds in specific regions of the galaxy, which could then have a direct impact on SF. Figure \ref{fig:extreme} showcases the spatial distributions of the top 100 most extreme clouds within the context of M51 for the different cloud properties considered: mass, average surface density, aspect ratio, and signatures of high-mass star formation. Table \ref{tab:extreme} holds the expected cloud fractions according to the global distribution of MCs across M51 (i.e. number of clouds in an environment divided by the total number of clouds), as well as the fractions reported for the tails of some of the analysed distributions (i.e. number of extreme clouds in an environment divided by size of extreme sub-sample, $N=100$). If the environment has no direct role in dictating the existence of such extreme clouds, we would expect the fractions for the extreme clouds to reflect the global cloud fractions. In the sections below we analyse each set of extreme clouds in more detail, and put those in context with the expected trends as per other literature results.

\begin{figure}
    \centering
    \includegraphics[width=0.4\textwidth]{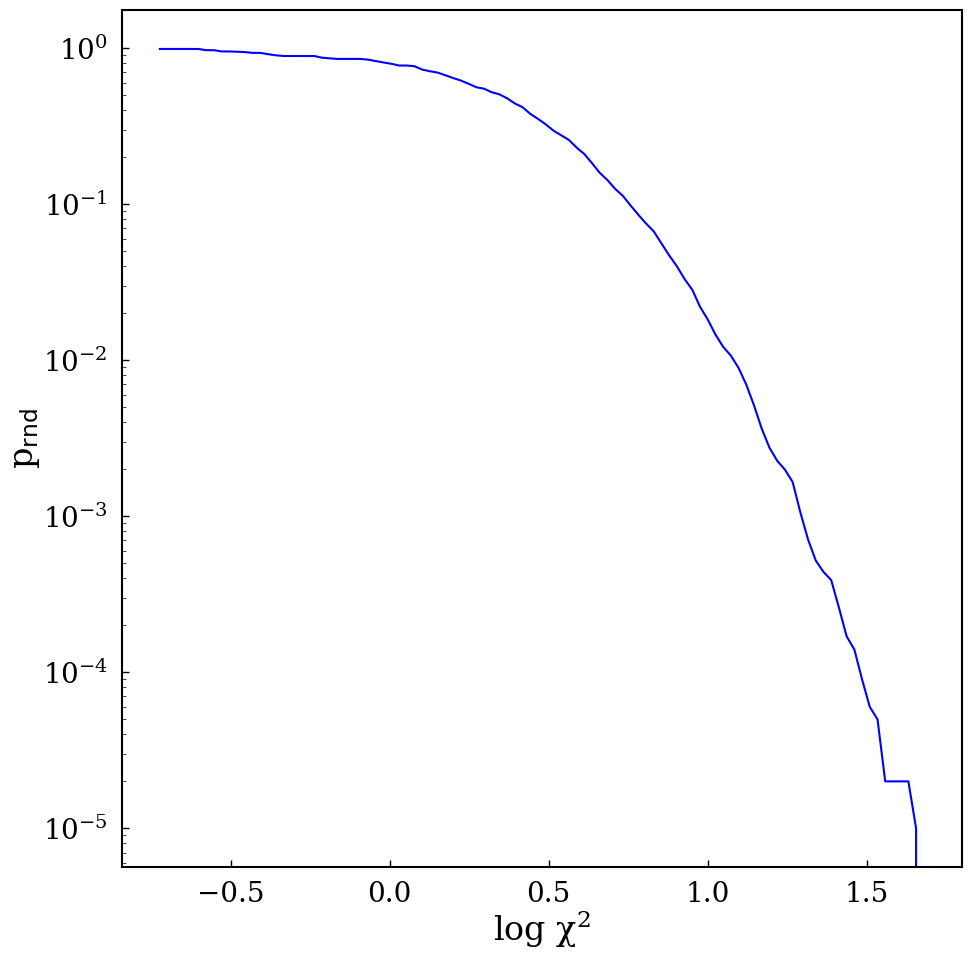}
    \caption{Cumulative distribution of the derived $\chi^2$ values for $100\,000$ random draws of $N=100$ clouds from our molecular sub-sample. Effectively, the y-axis represents the probability, $\text{p}_{\text{rnd}}$, of obtaining a given value of $\chi^2$ from a pure random draw.}
    \label{fig:chi2}
\end{figure}

To determine if the distribution of our extreme clouds is significant, we conduct a Pearson $\chi^2$ statistical analysis, which compares the observed distribution of a sample against a theoretical distribution and searches for similarities in frequencies. The $\chi^2$ value is given by the below expression:

\begin{equation}
    \chi^2 = \sum^{n}_{i=1} \frac{\left( O_i - E_i \right)^2}{E_i},
    \label{eqn:chi2}
\end{equation}

\noindent where $n$ is the number of environments considered (i.e. NB, MR, SA and IA), $O_i$ is the number of observed counts in environment $i$ (i.e. number of clouds), and $E_i$ is the number of expected counts within environment $i$ for a sample of size $N$, such that $E_i=f_i\,N$ with $f_i$ representing the probability of a cloud belonging to environment $i$ (i.e. the fraction of our science sample situated in each environment, listed in Table \ref{tab:extreme}). Here, we use our molecular sub-sample as our theoretical distribution, and calculate the $\chi^2$ statistics between our top 100 extreme clouds and the theoretical distribution. To test if the derived $\chi^2$ values are statistically significant, we determine the likelihood ($\text{p}_{\text{rnd}}$) of obtaining our calculated $\chi^2$ values if we randomly draw $N = 100$ clouds from our science sample (without replacement). To do so, we performed $100\,000$ random draws of $N = 100$ clouds, and determined the $\chi^2$ value for each draw (Eq. \ref{eqn:chi2}) against the expected or theoretical distribution (i.e. our science sample). We build a cumulative distribution of the $100\,000$ derived $\chi^2$ values to illustrate the likelihood of obtaining a certain $\chi^2$ value from pure random sampling, as shown in Fig. \ref{fig:chi2}. By comparing the $\chi^2$ values of our extreme sub-samples to the values resulting from random sampling, we are able to determine how likely we are to retrieve the observed extreme cloud distribution from a random sampling of the global population, and therefore judge whether any observed differences are statistically significant. In cases where the likelihood is low, the large-scale environment may have a direct role in promoting those specific types of extreme clouds. The exact values resulting from the $\chi^2$ statistics for our extreme clouds (listed in Table \ref{tab:extreme}) should not necessarily be taken at face value, and should serve instead as a means to compare between the different sub-sets of extreme clouds. From this analysis, we can see that properties like the footprint area ($A$), the medial axis length ($\text{L}_{\text{MA}}$), and aspect ratio have the highest $\text{p}_\text{rnd}$ values, suggesting that these properties mimic the general distribution more, while others like surface density have much lower $\text{p}_\text{rnd}$. In the following sections, we explore these trends in more detail, looking at extreme clouds in terms of their mass/surface density in §\ref{sec:massive}, elongation in §\ref{sec:elongated}, and high-mass star formation in §\ref{sec:hmsf}.

\subsubsection{Most massive/highest surface density}
\label{sec:massive}

Some observational studies of M51 have suggested that the spiral arms are the preferred location of the most massive MCs \citep[e.g.][]{koda_dynamically_2009, miyamoto_influence_2014, colombo_pdbi_2014} - a natural consequence of spiral arms hosting more material, which increases the frequency of cloud-cloud collisions leading to the formation of high-mass objects \citep[e.g.][]{dobbs_gmc_2008}. In the previous section (§\ref{sec:cmd}) it was already highlighted that the spiral arms seem able to form higher mass MCs than the inter-arms, despite having similar distributions in the low-to-intermediate mass range. It follows that when isolating the most massive MCs in our molecular sub-sample, the spiral arms boast a much higher number of these high-mass MCs than the inter-arms - a trend that does not follow the overall distribution of MCs across M51 and is therefore likely to be significant. Furthermore, a significant percentage of these extremely massive clouds reside in the molecular ring (a factor 4 more than what would be expected from statistics), a region also known to harbour an accumulation of material. The lack of high-mass MCs in the nuclear bar and inter-arms due to complex dynamics and shear was already seen and addressed in §\ref{sec:cmd}.

When looking at the bottom left panel of Fig.~\ref{fig:extreme}, it is clear that the densest MCs in our science sample prefer the spiral arms (an increase of roughly 63\% relative to the cloud fraction expected from the overall statistics). Additionally, these extremely dense clouds are heavily concentrated towards the inner regions of M51, again hinting at some strong radial trends (further analysed in §\ref{sec:radial}). Moreover, there is an increase of extremely dense clouds in the molecular ring relative to the expected statistics. From the figure, these dense MR clouds mostly correspond to the beginning of the spiral arms of M51 within the ring. The densest clouds seem to mostly be located in crowded areas where intense shear is absent, which hints at a dependence of the dense gas mass fraction as a function of a large-scale dynamical environment.   

For the most massive and highest surface density clouds we obtain $\chi^2$ values of 25.6 and 230, respectively, with corresponding likelihoods $\text{p}_{\text{rnd}}$ of $4\times10^{-4}$ and $<10^{-5}$. Both extreme sub-samples are therefore unlikely to be randomly drawn from our science sample, especially the highest surface density clouds. It is important to note that although a small amount of these extreme clouds have masses/surface densities that we do not necessarily trust due to saturation effects or observational limits (see Appendix \ref{app:mass_sd_unc}), the trends we report remain the same when removing these more uncertain clouds from our analysis.

\subsubsection{Most elongated}
\label{sec:elongated}

In their numerical study of GMCs of a two-armed spiral galaxy, \cite{duarte-cabral_what_2016} found that although the median properties of the inter-arm and spiral arm populations are similar in terms of aspect ratio, the most elongated MCs in their sample belong almost exclusively to the inter-arms. This could be suggestive of intense shear stretching massive MCs as they exit the spiral arms into the inter-arms \citep[e.g.][]{koda_dynamically_2009}, or disruption caused by stellar feedback \citep[e.g.][]{meidt_short_2015,chevance_lifecycle_2020}.

We have seen from §\ref{sec:mcproperties} that there are no significant differences in cloud elongation between the IA and SA populations when looking at the medians of either metric of aspect ratio. When looking at the top 100 most elongated MCs according to their $AR_{\text{MA}}$ instead, the majority of highly elongated clouds are located in the inter-arms. However, using $AR_{\text{a/b}}$ instead gives no significant increase of highly filamentary structures in the inter-arms. This discrepancy between the two metrics might be due to filamentary clouds that have a "curved" nature (e.g. ring-like), which would have a large $AR_{\text{MA}}$ but a low $AR_{\text{a/b}}$ (further discussed in Appendix \ref{sec:appB}). Clouds such as these might be potential "bubbles" which are driven by stellar feedback \citep[e.g.][]{watkins_phangsjwst_2023,barnes_phangsjwst_2023}. Both metrics report higher fractions of extremely elongated clouds in the nuclear bar than expected from statistics alone (factor 2.5 increase for $AR_{\text{MA}}$ and 5 for $AR_{\text{a/b}}$), reflective of the complex dynamical processes and intense shear seen towards that region. The molecular ring presents the most drastic difference between the two metrics (as was already pointed out in §\ref{sec:mcproperties}), with the moment aspect ratio metric reporting a significant increase of extremely elongated clouds whilst the medial axis aspect ratio sees no increase at all. We derive a $\chi^2$ value of 15.7 and $\text{p}_\text{rnd}=0.004$ for the extreme $\text{AR}_\text{MA}$ sub-sample, and $\chi^2=61.7$ and $\text{p}_\text{rnd}<10^{-5}$ for $\text{AR}_\text{a/b}$. The statistics suggest that these extreme sub-samples deviate from the theoretical distribution, however the deviations seem to be driven predominantly by the nuclear bar of M51, where both metrics agree on a surplus of extremely elongated clouds.

The discrepancies between $AR_{\text{a/b}}$ and $AR_{\text{MA}}$ and in particular, their different behaviour with different cloud morphologies (further discussed in Appendix \ref{sec:appB}) make it hard to draw any definite conclusions. In an attempt to isolate the truly elongated clouds, we instead retrieve the most elongated MCs from both metrics combined. To do so, we first standardise both distributions to make them comparable. We scale the $AR_{\text{MA}}$ and $AR_{\text{a/b}}$ distributions to both have a standard deviation of 1 and a mean of 0. Looking at the clouds with aspect ratio above $3\sigma$ in both rescaled distributions returns just 23 MCs - 9\% in NB, 9\% in MR, 30\% in SA and 52\% in IA. If we relax the threshold down to $2\sigma$, 88 MCs are considered and the percentages become 3\% in NB, 8\% in MR, 41\% in SA, and 48\% in IA, as shown in Table \ref{tab:extreme}. In either case, there is no significant increase of highly elongated MCs towards the inter-arms, but the amount of extremely elongated clouds in the nuclear bar remains statistically significant. The molecular ring population still hosts a significant fraction of these extreme clouds relative to the expected distribution. The MR is a region known to have low shear \citep[e.g.][]{meidt_gas_2013, querejeta_gravitational_2016}, so it could be that stellar feedback is the mechanism responsible for disrupting the MCs in this environment, although a more detailed cloud classification is needed to draw any definite conclusions. The distribution of our $AR_{\text{scaled}}$ sub-sample has a $\chi^2$ value of 15.3 and a likelihood $\text{p}_{\text{rnd}}$ of 0.005. We also do not observe any trend of cloud size (either through equivalent radius or medial axis length) across the large-scale environment.

\begin{figure}
    \centering
    \includegraphics[width=0.4\textwidth]{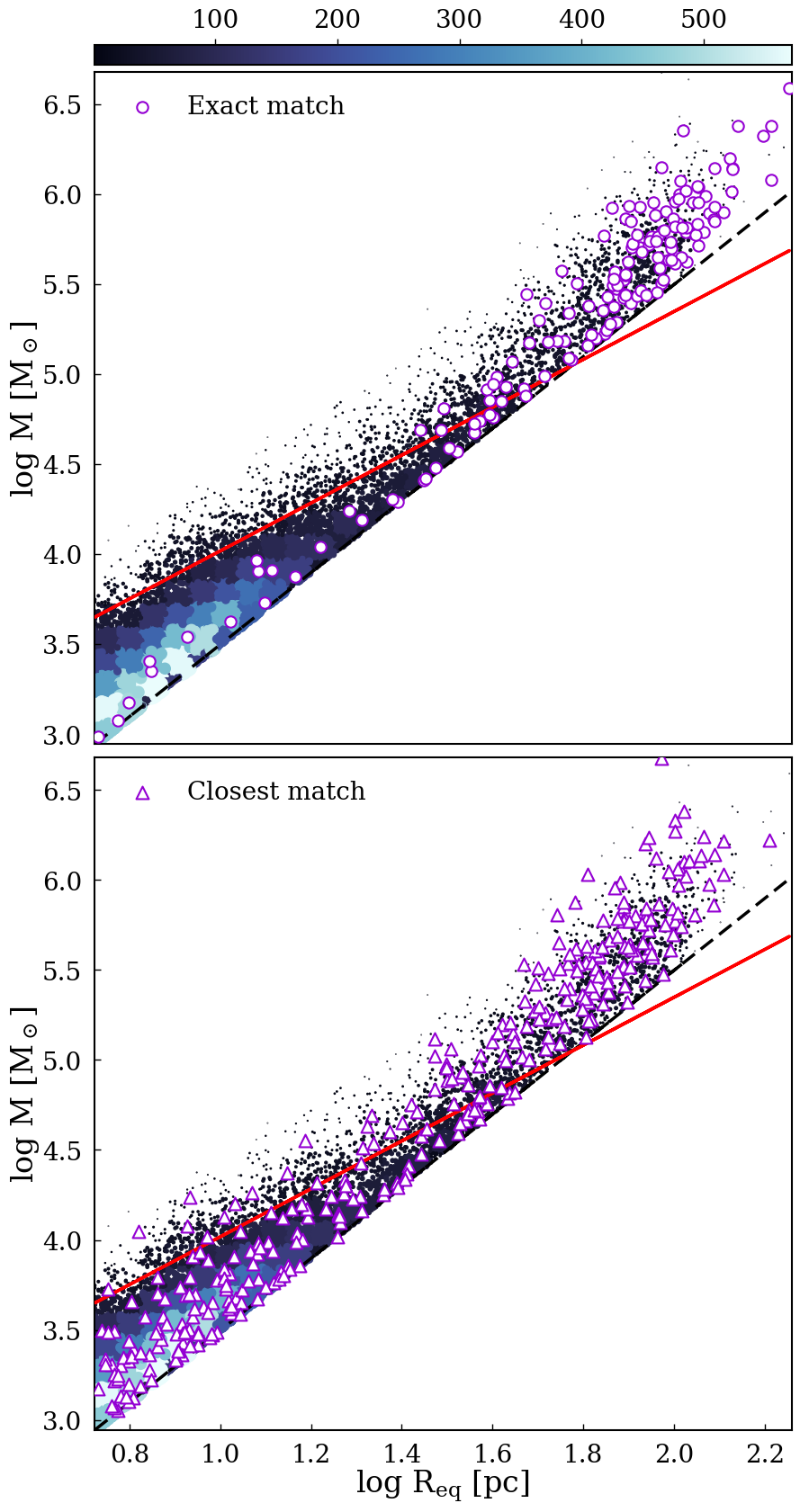}
    \caption{Mass-size relation for our science sample. On both panels, the blue scale indicates the density of points and the dashed black line the ${\Sigma = 10\,\sunpc}$ cut applied to obtain our high-fidelity sample. The continuous red line represents the empirical threshold for HMSF from \protect\cite{kauffmann_how_2010}. This threshold has been scaled from the original (given the different opacity laws used) to ${M = 487\,R^{1.33}}$, where the mass is in M$_{\odot}$ and the radius in pc. The violet dots on the top panel represent the clouds with an exact match to an 8\,$\upmu$m core from \protect\cite{elmegreen_highly_2019} (indicative of HMSF), whilst the violet triangles in the bottom panel depict the closest cloud match to a given source.}
    \label{fig:larson_sources}
\end{figure}

\subsubsection{High-mass star forming}
\label{sec:hmsf}

The highest mass and densest MCs in our sample are preferentially located in the spiral arms (and also the molecular ring), as shown in §\ref{sec:massive}. However, whether this enhancement of massive/dense clouds is then reflected on a different type of SF happening in those clouds, is still unclear. For instance, if high-mass star formation (HMSF) requires a cloud reaching higher masses or densities, then environments with a surplus of massive/dense MCs will also have a higher frequency of clouds hosting HMSF compared to the statistical distribution of clouds in general. In particular, if HMSF sites are enhanced in spiral arms then it may mean that SF is directly enhanced from the passage of the spiral density wave \citep[e.g.][]{roberts_large-scale_1969, lord_efficient_1990, seigar_test_2002}, rather than just a byproduct of orbit crowding in spiral arms \citep[e.g.][]{elmegreen_density_1986, foyle_arm_2010, ragan_prevalence_2016, urquhart_sedigism-atlasgal_2020}. 

We thus investigate the HMSF potential for our sample of clouds, by using the empirical relation derived by \cite{kauffmann_how_2010} to define a surface density threshold above which clouds are potential hosts for HMSF. The original HMSF threshold in \cite{kauffmann_how_2010}, $M$[M$_\odot$]$= 870 (R$[pc])$^{1.33}$, was determined with the opacity law $\kappa_{\lambda} = 12.1 \, (\lambda/250 \upmu \text{m})^{1.75}$ cm$^2$g$^{-1}$. In turn, our adopted opacity law is  $\kappa_{\lambda} = 21.6 \, (\lambda/250 \upmu \text{m})^{2}$ cm$^2$g$^{-1}$ from \cite{ossenkopf_dust_1994}, and thus we scale the HMSF threshold down to $M$[M$_\odot$]$= 487 (R$[pc])$^{1.33}$. The difference in dust mass from using either our specific opacity with a dust emissivity index of $\beta=2$ or the opacity employed by \cite{kauffmann_how_2010} with $\beta=1.75$ is only around 20\%, a small difference given the uncertainties on the masses themselves. 

Figure \ref{fig:larson_sources} displays the mass-size distribution of the clouds in our sample, with the solid red line representing the aforementioned HMSF threshold scaled to our adopted absorption coefficient. Around 15\% of our science sample sits above the HMSF threshold (2022 out of 13258 MCs). Of these 2022 MCs, 3\% belong to the nuclear bar, 6\% to the molecular ring, 53\% to the spiral arms, and the remaining 38\% to the inter-arms. The molecular ring and spiral arm fractions resulting from adopting this single threshold for HMSF are significantly higher than what would be expected from the overall distribution (2\% and 45\%, respectively, see $f$ in Table \ref{tab:extreme}). This indeed suggests that MCs in the molecular ring and spiral arms could be more prone to potentially host HMSF. 

In Fig.~\ref{fig:larson_sources} we also highlight known 8 $\upmu$m sources in M51 from \cite{elmegreen_highly_2019}, which are thought to trace the highly embedded and young stellar population of the galaxy. These 8\,$\upmu$m cores have a typical diameter of $3"$ (barely above the $2.4"$ FWHM resolution of the data), which corresponds to a physical size of about 110 pc for M51. Given the distances involved as well as the physical sizes of these sources (much larger than our typical cloud), it is likely that these are tracing unresolved sites of clustered HMSF. As such, we use these 8 $\upmu$m sources as HMSF signposts to determine the validity of an empirical surface density threshold. Using the catalogued central position of each 8 $\upmu$m source from \cite{elmegreen_highly_2019}, we create circular masks for each individual source with a $3"$ diameter. Cross-matching the source masks with the footprint masks of our MCs gives 509 matches out of 670. Over 100 sources are dismissed in this step: some fall outside the bounds of our map (the original catalogue includes NGC 5195), others are encompassed in diffuse clouds that are not considered in our molecular sub-sample, and others are not embedded anymore (i.e. young clusters also showing in the visible) leading us to not be able to measure any visual extinction in that region. Out of these 509 sources, 169 match with only 1 MC, whilst the remaining 340 match with multiple of our MCs. In order to perform an environmental analysis, we choose to only keep the match with the closest cloud (i.e. shortest distance between centroid of source and centroid of cloud). The top panel of Fig. \ref{fig:larson_sources} illustrates the exact source-cloud matches, whilst the bottom panel depicts the closest matches in the multiple clouds cases.

Our cross-matching results in 509 8 $\upmu$m sources from \cite{elmegreen_highly_2019} matching with 460 of our MCs (49 MCs have multiple associated sources, whilst the rest have unique, one-to-one matches), which are shown in Fig.~\ref{fig:larson_sources}. Out of these 460 MCs with an associated HMSF signpost, 279 are above the empirical HMSF line, whilst 181 are below. Adopting such a surface density threshold would cause us to miss roughly 65\% of true positives (i.e. MCs with an associated 8 $\upmu$m core yet are below the HSMF line). It is important to note that due to our source-cloud matching by proximity, some clouds may not be the true hosts of the $8 \, \upmu$m source, which will affect this fraction of missed true positives. Furthermore, while $8 \, \upmu$m can trace young clusters, it is not quite able to trace the younger and much more embedded young stellar objects present in the densest parts of MCs (i.e. tracers of "on-going" SF), and therefore our sample of HMSF signposts is by no means complete.

\begin{figure*}
    \centering
    \includegraphics[width = 0.9\textwidth]{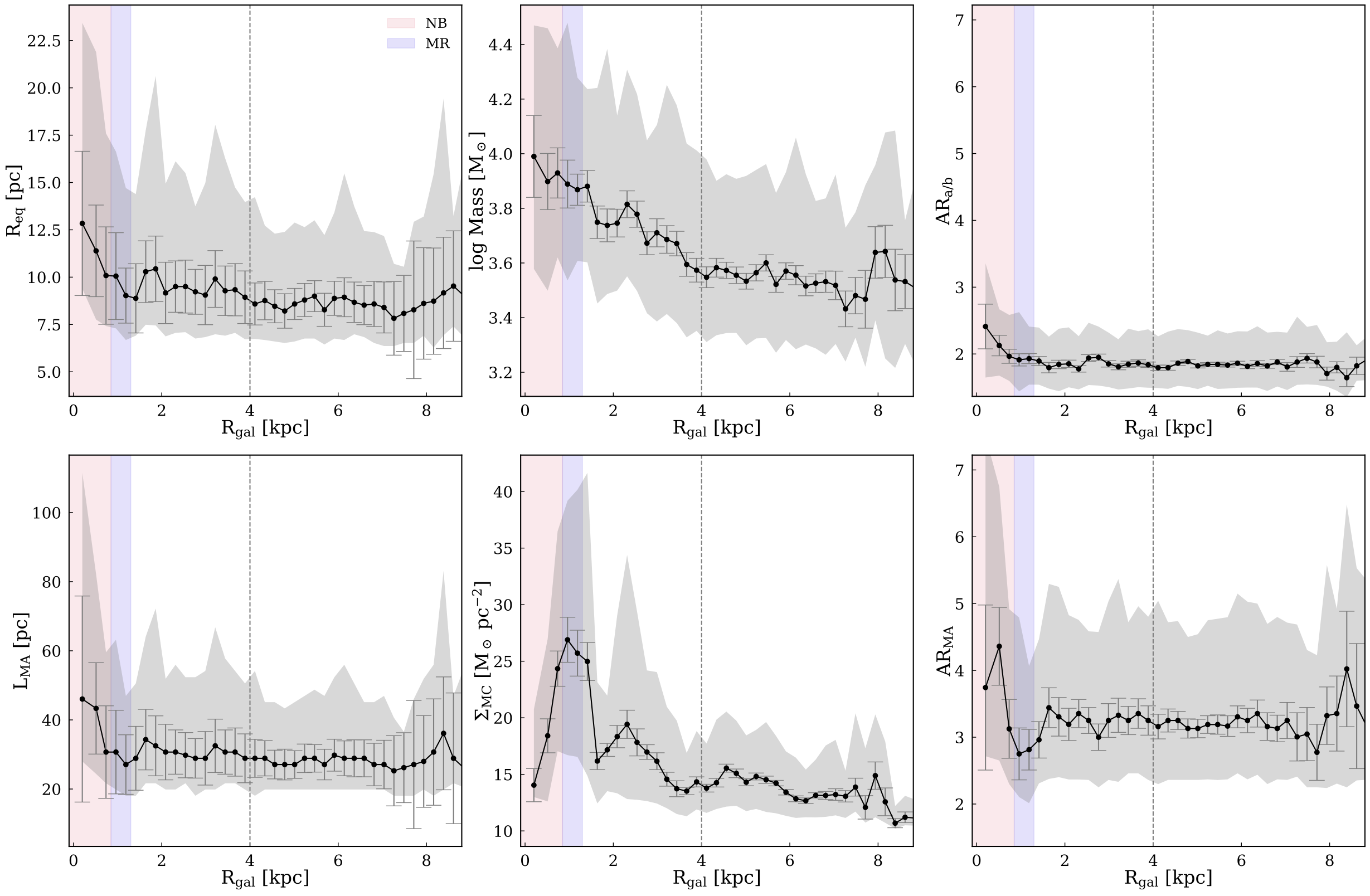}
    \caption{Some properties of our molecular clouds with galactocentric distance, $R_{\text{gal}}$: equivalent radius $R_{\text{eq}}$ (\textit{top left}), mass $M$ (\textit{top middle}), major/minor axis ratio $AR_{\text{a/b}}$ (\textit{top right}), medial axis length $L_{\text{MA}}$ (\textit{bottom left}), average surface density of MCs $\Sigma_{\text{MC}}$ (\textit{bottom middle}), and medial axis aspect ratio $AR_{\text{MA}}$ (\textit{bottom right}). For all panels, the running median of each respective property across the radial bins considered is portrayed by filled black circles and the solid black line connecting them. The grey-shaded region represents the corresponding interquartile range of the distribution. The grey error bars depict the standard error on the median ($1.253 \, \sigma / \sqrt{N}$, where $N$ is the bin count). The vertical dashed line is placed at $R_\text{gal}=4$\,kpc.}
    \label{fig:rgal_global}
\end{figure*}

Although there is a concentration of potential HMSF signposts towards the upper right corner of both panels of Fig.~\ref{fig:larson_sources} (i.e. towards higher-mass objects), there is still a significant amount of low-density and low-mass MCs that are HMSF candidates. In fact, of the highest surface density and highest mass clouds analysed in §\ref{sec:massive}, only 11 and 43, respectively, have an associated 8 $\upmu$m source. Additionally, there seems to be an increase of clouds with an associated 8\,$\upmu$m source towards the molecular ring and the spiral arms, as shown in the bottom right panel of Fig.~\ref{fig:extreme} and also by the $\chi^2$ and $\text{p}_\text{rnd}$ values we obtain (23.1 and $10^{-4}$, respectively), which we also noted from applying the \cite{kauffmann_how_2010} HMSF threshold. Despite this increase in HMSF signposts towards particular environments, from this analysis alone we are not able to distinguish between a higher star formation rate in more crowded regions (MR and SA) and an actual increase of star formation efficiency (i.e. the environment itself has a direct impact on the star formation process, rather than just gathering star-forming material). Even though our HMSF signpost sample is not complete, it does seem that there is a complex interplay of effects leading towards HMSF rather than a simple density/mass threshold from which all clouds can start forming massive stars. It is worth noting that the HMSF threshold proposed by \cite{kauffmann_how_2010} was originally derived for infrared dark clouds, which are very high column density objects. Our data is much more sensitive to the lower end of column density, and therefore applying this threshold may not be particularly relevant or useful. This analysis will benefit from higher resolution mid-IR observations (e.g. from JWST) that are able to probe a younger stellar population that is too embedded to show in $8\,\upmu$m with previous data for nearby galaxies.

\section{Trends with galactocentric radius}
\label{sec:radial}

In the previous sections we have looked at whether galactic environments have a direct impact on the characteristics of their cloud population and consequently SF, and found that although the large-scale dynamics do shape cloud characteristics, there is no strong sign that SF efficiency is enhanced towards any environment in particular \citep[see also][]{querejeta_stellar_2021}. Non-axisymmetries in the gravitational potential (i.e. spiral arms, nuclear bars) cause the gas in a galaxy to continuously flow not just between large-scale environments, but also radially. Naturally, we would expect the distribution of the ISM to be heavily influenced by these flows.  \cite{schuster_complete_2007}, for example, find a factor 20 decrease of molecular mass surface densities from the centre to the outskirts of M51 ($R_{\text{gal}} \sim 12$kpc). More recently, \cite{tres_simulations_2021} also identify a trend of decreasing cloud masses towards larger galactocentric radii in their simulated MC population of an M51-like galaxy. We thus make use of our high-resolution dataset to analyse the distribution of several properties of our MC sub-sample as a function of galactocentric radius. 

\subsection{Radial profiles}

Figure \ref{fig:rgal_global} shows the radial profiles of the MCs properties analysed in this paper, where M51 has been divided into 39 concentric bins of width 225 pc, with the exception of the first and last bin, which span 400 and $\sim$440 pc, respectively, given the lack of clouds seen at those radii. From the middle panels of the figure, we confirm that there is a general declining trend with galactocentric distance for both cloud mass and cloud average surface density, although the decline is less pronounced past $R_{\text{gal}}=4$\,kpc. The sudden spike in cloud masses at around $R_{\text{gal}}=8$\,kpc seems to be mostly due to a large group of MCs concentrated towards the end of the spiral arm leading up to NGC 5195. There is no obvious radial trend of cloud size either through equivalent radius or medial axis length (leftmost panels of Fig.~\ref{fig:rgal_global}), except in the first few bins corresponding to the nuclear bar, where clouds seem to be longer. Both metrics of aspect ratio (rightmost panels of Fig.~\ref{fig:rgal_global}) remain fairly constant at all radii, apart from a slight increase for the first radial bins again corresponding to the nuclear bar. 

\subsection{Radial profiles per large-scale environment}

\begin{figure}
    \centering
    \includegraphics[width=0.4\textwidth]{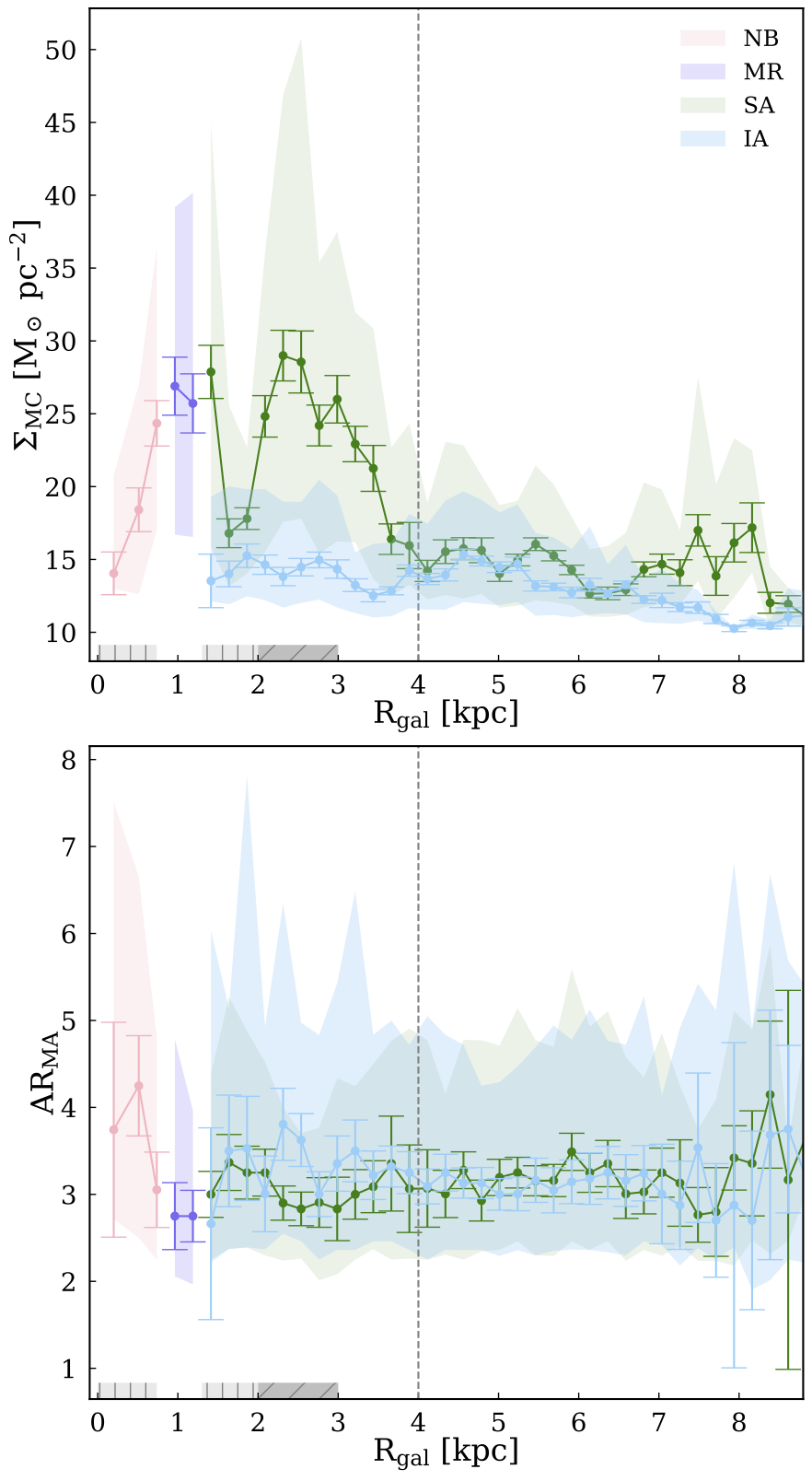}
    \caption{Average cloud surface density, $\Sigma_{\text{MC}}$ (\textit{top}), and medial axis aspect ratio, $AR_{\text{MA}}$ (\textit{bottom}) across galactocentric distance for the different dynamical environments of M51: nuclear bar (NB in pink), molecular ring (MR in purple), spiral arms (SA in green) and inter-arms (IA in light blue). The coloured circles and lines represent the running median of the relevant property. The shaded regions are the corresponding interquartile range of the distributions. The coloured errorbars illustrate the respective standard error on the median ($1.253 \sigma / \sqrt{N}$). The vertical dashed line is placed at $R_\text{gal}=4$\,kpc. The light grey shaded region with vertical hatches adjacent to the x-axis represents known regions of little to no SF in M51 \citep[e.g.][]{schinnerer_pdbi_2013}. The darker grey shaded region with diagonal hatches represents a region with intense SF \citep[e.g.][]{meidt_gas_2013}.}
    \label{fig:rgal_env}
\end{figure}

Simple 1D radial profiles average different environments together; looking instead at the same radial bins but within each environment separately will highlight any interesting signatures that might otherwise get washed out by the mixing with other environments. Figure\,\ref{fig:rgal_env} illustrates the average cloud surface density and medial axis aspect ratio for the separate galactic environments of M51 with galactocentric distance. The remaining properties from Fig.~\ref{fig:rgal_global} do not show significant changes, apart from cloud mass which has a similar trend to $\Sigma_{\text{MC}}$.   

The $\Sigma_{\text{MC}}$ radial profiles of the separate large-scale environments have very distinct features (shown in the top panel of Fig.~\ref{fig:rgal_env}). As can be seen from the figure, towards the inner galaxy there is a sudden drop of $\Sigma_{\text{MC}}$ at $\sim$1.7 kpc in the spiral arms (also present when building a radial profile of each pixel's surface density within our mask of the spiral arms), and it coincides with a known region of little to no SF \citep[e.g.][]{meidt_gas_2013, querejeta_dense_2019}. In their kinematic study of M51 using PAWS data, \cite{colombo_pdbi_2014_moment} find inflowing non-circular motions driven by the start of the spiral arms between $1.3 < R_{\text{gal}} < 2$ kpc, coinciding with our dip in $\Sigma_{\text{MC}}$ for the spiral arms. Additionally, \cite{henry_star_2003} find a deviation from a pure spiral pattern caused by two dominant arms ($m = 2$ mode) for $1 < R_{\text{gal}} < 2.2$ kpc \citep[also seen by ][]{colombo_pdbi_2014_moment}, which could increase the streaming motions of the gas, depleting the available reservoir at those radii and lowering the observed densities. Once reaching the molecular ring, cloud surface densities in the spiral arms seem to rise again, likely from gas being stalled against the MR dynamical barrier. There is also little to no SF detected for the inner $\sim 750$ pc of M51, where peculiar motions driven by the bar are dominant and heavily disrupt and disperse the gas \citep[e.g.][]{colombo_pdbi_2014_moment}.

\begin{figure}
    \centering
    \includegraphics[width=0.4\textwidth]{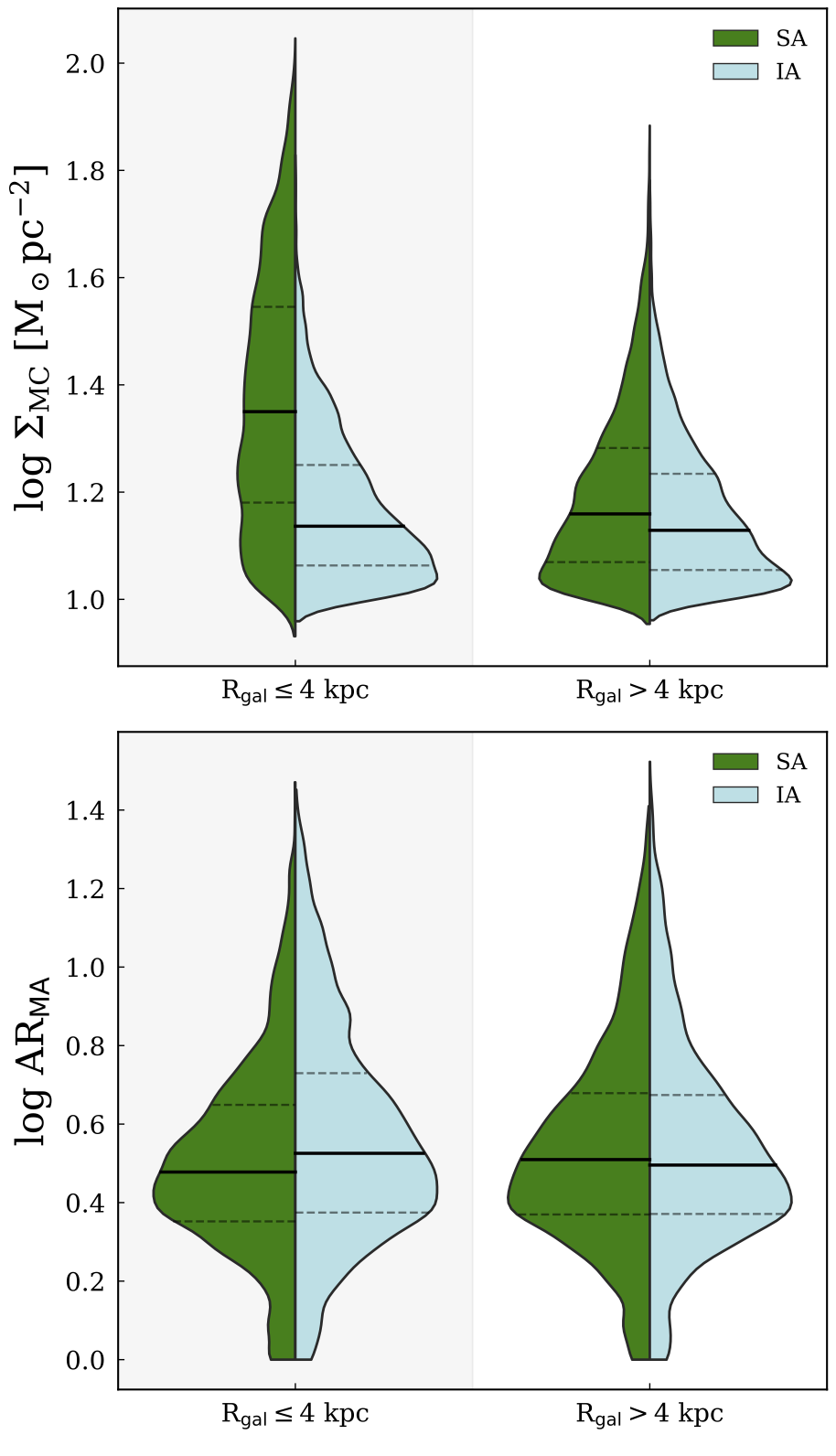}
    \caption{Violin plots showing the contrast between average cloud surface density ($\Sigma_{\text{MC}}$, \textit{top}) and medial axis aspect ratio ($\text{AR}_{\text{MA}}$, \textit{bottom}) of MCs in the spiral arms (SA in green) and in the inter-arms (IA in blue). For both panels, the cloud populations of each environment are shown for the inner galaxy on the left (shaded region, ${\text{R}_{\text{gal}}\lesssim4\,\text{kpc}}$) and for the outer galaxy on the right (${\text{R}_{\text{gal}}>4\,\text{kpc}}$). For all violin plots the solid black line represents the median of the distribution, whilst the dashed black lines indicate the upper and lower quartile (as seen from top to bottom).}
    \label{fig:violin}
\end{figure}

As shown in the top panel of Fig.~\ref{fig:rgal_env}, the distribution of $\Sigma_{\text{MC}}$ for the inter-arms is fairly constant across radial distance, meaning that the declining trend seen for the global profile is indeed driven by the bar and spiral arms of M51. Additionally, the tentative flattening of cloud densities past ${R_{\text{gal}}\sim4}$ kpc witnessed in Fig.~\ref{fig:rgal_global} is much more pronounced when looking at the spiral arms, and the phenomena causing it does not seem to affect the inter-arms. To further investigate this shift in behaviour, Figure \ref{fig:violin} highlights the differences in properties of the cloud populations in the inner ($R_\text{gal}<4$\,kpc) and outer ($R_\text{gal}>4$\,kpc) galaxy, for both the spiral arms and the inter-arm regions. As was already seen in the radial profiles, it is clear that MCs in the inner spiral arms are much denser than IA clouds at the same radii, whilst the average density of both populations is similar at larger galactocentric radii (top panel of Fig. \ref{fig:violin}). The same trend is seen for cloud mass, although less pronounced. The most elongated clouds in the inner galaxy seem to develop in the inter-arms (since the upper part of the inner IA violin plot is more populated, shown in bottom panel of Fig. \ref{fig:violin}), whilst at larger radii the SA and IA distributions are virtually identical.

The clue to this behaviour may lie in the nature of the spiral arms of M51. If M51 was composed of a single quasi-stationary density-wave with a fixed pattern speed, we would expect to see enhanced surface densities/masses throughout the entire spiral arms (relative to the inter-arm regions), since the gas would be harboured and compressed in the strong spiral gravitational potential well generated by the density wave \citep[for a review see ][]{lin_spiral_1964,binney_galactic_1987}. This behaviour is indeed similar to what we see in the top panel of Fig. \ref{fig:rgal_env} for $R_\text{gal}<4$\,kpc, but not so much for the outskirts of the galaxy. For a density-wave type of pattern, we would also expect to observe newborn stars within the spiral arms and increasingly older stars as you move along in azimuth (i.e. a stellar age gradient), which again is observed in M51 by some studies \citep[e.g.][]{abdeen_evidence_2022}, but not by others \citep[e.g.][]{schinnerer_pdbi_2017,shabani_search_2018}. In fact, several studies, both numerical and observational, argue against a fixed pattern speed in M51, and thus a single density-wave type of pattern \citep[e.g.][]{tully_kinematics_1974,elmegreen_spiral-arm_1989,meidt_radial_2008,dobbs_simulations_2010}. Instead, the spiral structure of M51 seems to have a more transient nature, which evolves dynamically with time as a function of the tidal interaction with its companion NGC 5195 \citep[e.g.][]{toomre_galactic_1972,elmegreen_spiral-arm_1989,dobbs_simulations_2010}. 

The top panel of Fig. \ref{fig:rgal_env} suggests that the gas in the spiral arms of M51 has two distinct behaviours. In the inner galaxy ($R_\text{gal}<4$\,kpc), the spiral arms boast much higher average cloud surface densities relative to the inter-arm regions, similar to the expected behaviour driven by a density-wave type of pattern which promotes a higher frequency of massive SA MCs. On the other hand, in the outer galaxy ($R_\text{gal}>4$\,kpc), cloud surface densities are very similar for both SA and IA. This change in behaviour occurs at around the same radii for which \cite{querejeta_gravitational_2016} and \cite{zhang_galaxy_2012} find significant changes in torque signs (at $3.8$\,kpc) and potential-density phase shifts (at $4.1$\,kpc), respectively, which the authors attribute to a co-rotation of the spiral pattern with the gas. Given that there is substantial evidence that M51 does not have a single pattern speed (as mentioned above), the notion of co-rotation becomes more complex; still it is clear that there is a sharp change in behaviour at this radius. It seems that, even though the spiral pattern is not rotating at a fixed speed in the inner galaxy, the gas is still rotating faster than the spiral arms, meaning that the gas feels the compression due to the passage through the spiral arm as it would on a density-wave type of pattern. As mentioned above, at large galactocentric radii ($R_\text{gal}>4$\,kpc) the SA cloud surface densities become more comparable to the inter-arms, suggesting that past $R_\text{gal}=4$\,kpc the spiral pattern and the gas are nearly co-moving. In other words, the outer spiral arms seem to be generated by local gravitational instabilities and behave more like material arms rather than a density-wave \citep[see also][]{miyamoto_influence_2014,colombo_pdbi_2014_moment}, which is likely due to the influence of the tidal interaction. The outer spiral arms in M51 are therefore unable to drive the same density enhancement seen in the inner arms \citep[e.g.][]{dobbs_simulations_2008}, since it seems that at large $R_\text{gal}$ the gas does not have enough time to cross the bottom of the spiral potential well given both the larger gas crossing times between the arms in the outskirts of the galaxy and the fact that the outer spiral arms seem to evolve at a much quicker rate relative to the inner spiral arms. Additionally, due to the weaker gravitational potential, the outer arms are less protected against shear thus resulting in their "fractured" appearance (as can be seen from the environmental mask in Fig.\,\ref{fig:env_mask}). In the shear-dominated inter-arm regions, we would not expect the gas to be much affected by the tidal interaction. We thus hypothesise that the sharp change in behaviour for the spiral arms at $R_\text{gal}=4$\,kpc is due to the dynamics of the interaction of M51 with its companion.

Additionally, in M51 SF occurs mostly on the convex side of the spiral arms at $2 < R_{\text{gal}} < 3$ kpc \citep[e.g.][]{meidt_gas_2013}, where we also find a peak in the average cloud surface density. The surrounding areas, namely within the inter-arm region, are likely to be affected by the feedback from these SF events, potentially leading to cloud disruption which could result in higher aspect ratios. For this region, there is a clear difference in the SA and IA medial aspect ratio profiles shown in the bottom panel of Fig.~\ref{fig:rgal_env}, where MCs in the inter-arms have higher aspect ratios than their counterparts in the SA. This could be consistent with stellar feedback disrupting the IA MCs, but could also be attributed to the strong shearing motions at this radii splitting clouds apart \citep[e.g.][]{dobbs_exciting_2013, miyamoto_influence_2014}. Furthermore, the shaded areas in the bottom panel of Fig.~\ref{fig:rgal_env}, which represent the interquartile range of cloud aspect ratios, seem to have different peaks depending on galactocentric radii. In the outer galaxy, clouds with high aspect ratios appear to be evenly distributed between SA and IA, but this is not the case at smaller galactocentric radii. For $R_\text{gal}<4$\,kpc, the majority of the highly elongated clouds seem to reside in the inter-arms, meaning that at these radii the inter-arms are more prone to develop the most elongated structures within our sample. This finding agrees well with the previously presented framework \citep[and with the work of ][]{duarte-cabral_what_2016}: in the inner galaxy where the pattern resembles a density-wave, the stronger spiral potential will protect clouds from intense shear within the arms but not in the inter-arm regions, leading to a higher frequency of fragmented/stretched clouds in the IA. This also explains why we do not find a surplus of extremely elongated IA clouds in §\ref{sec:elongated}, since we take the top 100 elongated clouds over the entire sample, effectively losing any effect the different spiral patterns may have on the clouds at different galactocentric radii.

%This conclusion, of course, also does not take into account the potential effects of stellar feedback mechanisms (as discussed in §\ref{sec:mcproperties}), although \cite{meidt_short_2015} suggest that shear, rather than feedback, seems to be the main driver of cloud disruption at small galactocentric radii.

To draw any firm conclusions, a more rigorous analysis in quantifying the shear and feedback in these regions is needed, as well as a more robust classification of truly filamentary clouds (as previously discussed in §\ref{sec:mcproperties}). This will be the focus of future work.

\section{Summary and conclusions}
\label{sec:sum_conc}

%The last numbered section should briefly summarise what has been done, and describe the final conclusions which the authors draw from their work.

In \cite{faustino_vieira_high-resolution_2023} we presented a new high-resolution extinction mapping technique, with which we mapped the gas content of M51 (NGC 5194) at a spatial resolution of $0.14"$ ($\sim 5$\,pc). Here, we extract clouds from our gas surface density map using \texttt{SCIMES} \citep[Spectral Clustering for Interstellar Molecular Emission Segmentation,][]{colombo_graph-based_2015,colombo_integrated_2019}. We compile a catalogue for all the identified clouds in M51 with measurements of several physical properties, which we release with this paper alongside all the footprint masks for each structure. With that catalogue we then analyse the sub-sample of molecular clouds across the galaxy, in search of any evidence of how their properties might be affected by large-scale galactic environment as well as a function of galactocentric radius (and the combination of the two). Our findings can be summarised as follows:

\begin{itemize}
    \item We find that molecular clouds residing in the centre of M51 show distinct differences from the disc population. Average cloud sizes, masses, surface densities, and aspect ratios (mostly within the nuclear bar) are higher in the inner few kiloparsecs of M51 than for the disc.
    \item We fit truncated power laws to the cumulative cloud mass distribution within each large-scale dynamical environment of M51. We find that the gas in M51 is preferentially organised into low-mass clouds in the disc and high-mass clouds in the centre. Additionally, the spiral arms and molecular ring host the highest concentration of high-mass clouds, whilst the inter-arms and nuclear bar distributions show a sharper decline towards higher masses.
    \item We isolate the most extreme clouds in our science sample with the purpose of ascertaining if a given cloud property is particularly enhanced towards a specific environment within the galaxy. We find no obvious enhancement of extremely large clouds (in both area and length) in any large-scale environment. On the other hand, there is a surplus of extremely elongated clouds in the nuclear bar region of M51. Additionally, the most massive and highest surface density clouds in our science sample show a clear preference for the molecular ring and spiral arms, suggesting that these environments host beneficial conditions for cloud growth.
    \item Although we detect an increase of high-mass star formation (as traced by 8 $\upmu$m from \citealt{elmegreen_highly_2019}) towards the spiral arms and molecular ring of M51, we are not able to determine if the higher star formation rate is simply due to crowding or an actual increase of star formation efficiency. We also find that assuming a surface density-mass threshold as an indicator of the ability of a given cloud to form stars appears to be an oversimplified approach that does not capture the complicated juxtaposition of effects in play. Although the SF analysis performed in this paper is very simplified, it nonetheless seems to agree with more in-depth star formation rates/efficiency studies, which find little evidence for enhanced star formation efficiencies in spiral arms \citep[e.g.][]{dobbs_properties_2011,urquhart_sedigism-atlasgal_2020,querejeta_stellar_2021}.
    \item There is no apparent trend between the galactocentric radius and cloud elongation or size for the disc of M51 when considering the entire population of clouds (without splitting into environments). There is a declining trend of surface densities towards the outskirts, as well as cloud mass and average cloud surface density.
    \item When using the 2D positional information to analyse the properties of clouds as a function of galactocentric distance for each environment separately, we find that although the average surface densities of the inter-arm molecular cloud population remain constant with galactocentric radius, the spiral arm clouds show a different behaviour at small and large radii. In fact, for $R_\text{gal}<4$\,kpc, there is a clear contrast between cloud surface densities of the inter-arms and spiral arms, whilst at larger radii they have similar radial profiles. Additionally, at small $R_\text{gal}$, the most elongated (i.e. highest aspect ratio) clouds seem to mostly belong to the inter-arms. 
    \item We find a sudden dip in surface densities at roughly 1.7 kpc in the spiral arms, where \cite{colombo_pdbi_2014_moment} detect an increase of non-circular motions driven by the start of the spiral arms and a potential perturbation in the spiral pattern \citep[see also][]{henry_star_2003}. For this radial region, we also observe higher cloud aspect ratios in the inter-arms than in the spiral arms.
\end{itemize}

Non-axisymmetric features (i.e. stellar bar, spiral arms) in M51 exert a substantial influence on how the gas is organised across the galaxy. There is a clear difference in characteristics between the cloud populations of the centre and the disc of M51. Peculiar motions driven by the nuclear bar heavily disrupt the clouds in that region, preventing and/or destroying higher mass objects and stretching out clouds, reflecting into high aspect ratios. Similarly, shearing motions (driven by the differential rotation of the gas) seem to have a similar effect in the inter-arms, albeit the observed characteristics of the inter-arm clouds could also be caused by stellar feedback. A more reliable quantification of cloud morphology is needed in order to distinguish the linearly elongated clouds driven by shear from the more distorted/ring-like clouds potentially associated with feedback regions. Nonetheless, in environments where shear is low (i.e. molecular ring and spiral arms), gas is allowed to accumulate resulting in the development of higher mass/density clouds. 

Additionally, we find that the tidal interaction between M51 and its companion has a strong influence on the cloud population of the spiral arms, but a minimal effect (if any) in the inter-arms clouds. At small radii, the spiral pattern resembles a density-wave type of pattern, where the strong spiral potential piles material up, and increased cloud-cloud collisions drive cloud masses up in the arms. Consequently, MCs in the inner spiral arms show enhanced surface densities/masses relative to their counterparts in the inter-arm regions. At large radii, where the tidal interaction seems to have a stronger influence, the spiral arms are evolving on a much shorter time-frame and appear to be driven by local gravitational instabilities, which affects both the gas and the stars similarly. Consequently, the outer spiral arms are not as able to promote cloud growth, resulting in the similarities seen between the inter-arm and spiral arm molecular cloud populations at those radii. 

This study demonstrates the power of larger number statistics on resolved cloud populations, as well as wider coverage across entire galaxies, in unravelling the potential effects of the environment on the formation and evolution of clouds. The spatially resolved information we obtain from our extinction-derived gas surface densities \citep{faustino_vieira_high-resolution_2023} allows for cloud-scale studies to be conducted across not only various galactic environments, but also across different galaxy types. Such exercises are fruitful in developing our understanding of SF as a galactic-driven process, and learn which mechanisms hinder or enhance the formation of stars (and where this occurs), which naturally has repercussions in the evolution of galaxies.

\section*{Acknowledgements}

%The Acknowledgements section is not numbered. Here you can thank helpful colleagues, acknowledge funding agencies, telescopes and facilities used etc. Try to keep it short.

We thank the anonymous referee for their comments and suggestions, which have helped improve the manuscript. HFV and ADC acknowledge the support from the Royal Society University Research Fellowship URF/R1/191609. TAD, NP, MWLS and MA acknowledge support from the UK Science and Technology Facilities Council through grants ST/S00033X/1 and ST/W000830/1. The calculations performed here made use of the computing resources provided by the Royal Society Research Grant RG150741. MQ acknowledges support from the Spanish grant PID2019-106027GA-C44, funded by MCIN/AEI/10.13039/501100011033. HFV acknowledges Sharon Meidt for the use of the PAWS environmental mask. Based on observations made with the NASA/ESA Hubble Space Telescope, which is operated by the Association of Universities for Research in Astronomy, Inc. (Program \#10452). DustPedia is a collaborative focused research project supported by the European Union under the Seventh Framework Programme (2007-2013) call (proposal no. 606847). The participating institutions are: Cardiff University, UK; National Observatory of Athens, Greece; Ghent University, Belgium; Université Paris Sud, France; National Institute for Astrophysics, Italy and CEA, France.

\section*{Data availability}
With this paper, we release the full catalogue off all clouds extracted with \texttt{SCIMES} and the respective cloud masks in {\url{https://dx.doi.org/10.11570/23.0030}}, as well as in the FFOGG (Following the Flow of Gas in Galaxies) project website ({\url{https://ffogg.github.io/}}).

%%%%%%%%%%%%%%%%%%%%%%%%%%%%%%%%%%%%%%%%%%%%%%%%%%



\bibliographystyle{mnras}
\bibliography{references} % if your bibtex file is called example.bib

%\begin{thebibliography}{99}
%\end{thebibliography}

%%%%%%%%%%%%%%%%%%%%%%%%%%%%%%%%%%%%%%%%%%%%%%%%%%

%%%%%%%%%%%%%%%%% APPENDICES %%%%%%%%%%%%%%%%%%%%%

\appendix

\section{Cluster properties and catalogue}
\label{sec:appA}

Alongside this paper, we make available the complete catalogue\footnote{\url{https://ffogg.github.io/ffogg.html}} of all the clouds extracted from our high-resolution extinction map of M51 using \texttt{SCIMES} and \texttt{ASTRODENDRO} (description of cluster extraction given in §\ref{sec:extraction}). Table \ref{tab:catalogue} specifies all the cluster properties contained in our catalogue.

\begin{table*}
    \begin{tabular}{ l c | p{13.5cm} }
    \hline
    \hline
    Catalogue Column & Variable & Description \\
    \hline
    \textit{ID} & & Unique ID number of cloud \\
    \textit{RA (J2000)} & & Right ascension of cloud in \textit{hh mm ss.ss} format \\
    \textit{Dec (J2000)} & & Declination of cloud in \textit{dd mm ss.ss} format \\
    \textit{RA\_deg} & & Right ascension of cloud (degrees) \\
    \textit{Dec\_deg} & & Declination of cloud (degrees) \\
    \textit{R\_gal} & $R_{\text{gal}}$ & Distance of cloud centre to the galactic centre (kpc) \\
    \textit{Sigma\_tot} & & Total sum of the gas mass surface density of every pixel in the cloud ($10^3 \, \sunpc$) \\
    \textit{Sigma\_avg} & $\Sigma_{\text{MC}}$ & Average gas mass surface density of cloud ($\sunpc$) \\
    \textit{Sigma\_peak} & & Peak gas surface density of cloud ($\sunpc$) \\
    \textit{Area\_ellipse} & & Area of the ellipse defined by the second moments of the cloud (pc$^{2}$) \\
    \textit{Area\_exact} & $A$ & Exact area of cloud (pc$^{2}$) \\
    \textit{R\_eq} & $R_{\text{eq}}$ & Equivalent radius estimated using the cloud's exact area (pc) \\
    \textit{Mass} & $M$ & Mass of cloud (M$_\odot$) \\
    \textit{Major\_axis\_a} & $a$ & Semi-major axis (pc) \\
    \textit{Minor\_axis\_b} & $b$ & Semi-minor axis (pc) \\
    \textit{AR\_ab} & $AR_{\text{a/b}}$ & Aspect ratio between semi-major and semi-minor axis \\
    \textit{PA} & & Position angle of cloud major axis, measured counter-clockwise from $+x$ axis (degrees) \\
    \textit{Length\_MA} & $L_{\text{MA}}$ & Length of the geometrical medial axis (pc) \\
    \textit{Width\_MA} & $W_{\text{MA}}$ & Width of the geometrical medial axis (pc) \\
    \textit{AR\_MA} & $AR_{\text{MA}}$ & Aspect ratio between the medial axis length and width \\
    \textit{Sat\_pix\_area} & & Portion of cloud's exact area that feature saturated/uncertain pixels (\%) \\
    \textit{Rel\_err} & $\sigma_\tau / \tau_V$ & Relative uncertainty on the cloud's opacity (and thus surface density/mass) from the dust extinction technique alone \\
    \textit{Env} & & Tag identifying the environment of the cloud  (NB=nuclear bar, MR=molecular ring, SA=spiral arms, IA=inter-arms) \\
    \textit{Robust\_bg} & & Tag identifying clouds detected in robust stellar backgrounds (1=robust, 0=faint) \\
    \textit{Molecular\_cut} & & Tag identifying predominantly molecular clouds, i.e. $\Sigma_{\text{avg}} > 10 \, \sunpc$ (1=molecular, 0=atomic) \\
    \textit{Size\_cut} & & Tag identifying clouds that pass our size criteria, i.e. $A > 3$ resolution elements (1=yes, 0=no) \\
    \hline
    \end{tabular}
    \caption{Description of the contents of the molecular cloud catalogue obtained from applying our high-resolution extinction-mapping technique to M51. The second column shows what the relevant quantities are referred as in this paper.}
    \label{tab:catalogue}
\end{table*}

\begin{figure}
    \centering
    \includegraphics[width=0.4\textwidth]{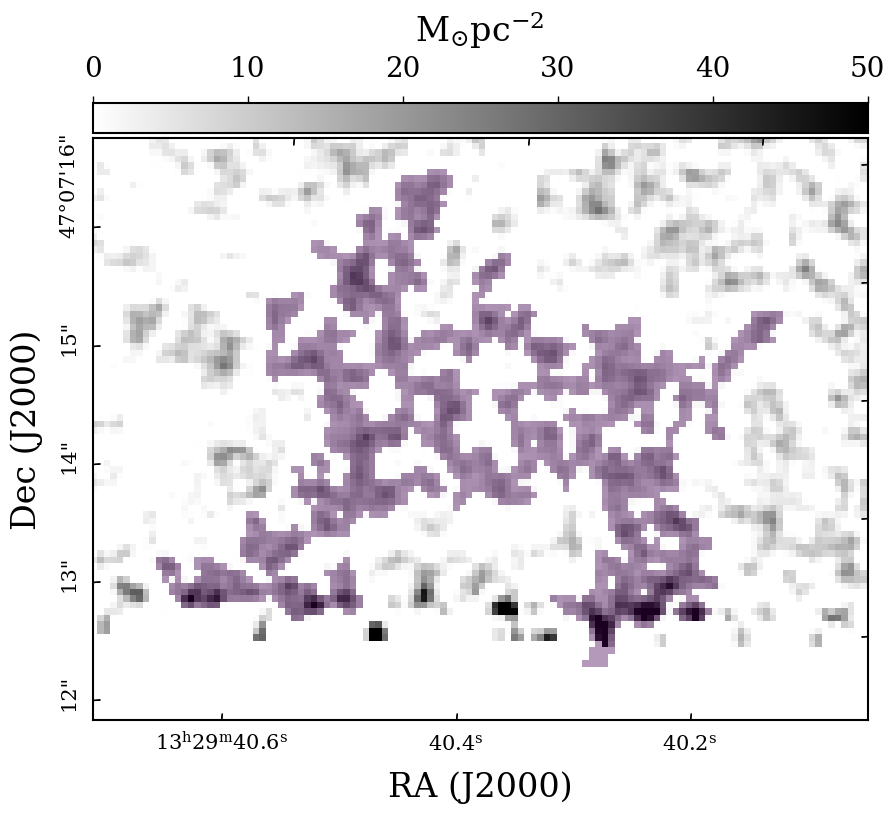}
    \caption{Footprint mask of a cloud in our sample (ID: 3, in purple) that satisfies our size and molecular criteria (i.e. \textit{Size\_cut}=1 and \textit{Molecular\_cut}=1), but is not set against a sufficiently robust stellar distribution (i.e. \textit{Robust\_cut}=0). Our gas surface density map is in the background greyscale.}
    \label{fig:I0_cut_mc}
\end{figure}

\subsection{Coordinates}

The right ascension and declination of each cloud's centroid (\textit{RA\_deg} and \textit{Dec\_deg}, respectively) was estimated by \texttt{ASTRODENDRO} when building the full dendrogram of our map. The galactocentric distance, \textit{R\_gal}, is estimated between the centroid position of each cloud and the centre of the galaxy. The galaxy's centre position is determined from the PAWS environmental mask. \textit{R\_gal} already takes into account M51's position angle and inclination \citep[173$^\circ$ and $22^\circ$, respectively, from][]{colombo_pdbi_2014_moment}.

\subsection{Geometrical properties}

From our full dendrogram, \texttt{ASTRODENDRO} also computes the area of the ellipse encompassing each cloud (\textit{Area\_ellipse}), the exact footprint area of a cloud (\textit{Area\_exact}), the semi-major and semi-minor axis of a cloud (\textit{Major\_axis\_a} and \textit{Minor\_axis\_b}, respectively), and the cloud's position angle (\textit{PA}, measured counter-clockwise in degrees from the +x axis in pixel coordinates). Using the exact footprint area of each cloud we compute its equivalent radius, \textit{R\_eq}, which is calculated assuming that the cloud is a circle such that $R_{\text{eq}} = \sqrt{A / \pi}$, where $A$ is the exact area of the cloud. 

Determining the aspect ratio of a cloud gives a basic estimate of the cloud's morphology: MCs with aspect ratio close to unity are circular, and MCs with high aspect ratio are elongated. The first aspect ratio we consider is the intensity-weighted moment aspect ratio, \textit{AR\_ab}, defined as the ratio between a cloud's semi-major axis (\textit{Major\_axis\_a}) and semi-minor axis (\textit{Minor\_axis\_b}). The other aspect ratio metric we use is the medial axis aspect ratio, \textit{AR\_MA}. The medial axis is the longest running spine of a cloud's mask that is also the furthest away from the external edges of the cloud (all holes within a cloud are filled before the calculation). It is not weighted by intensity, and is a purely geometrical approach. The medial axis is found by extracting the "skeleton" of the cloud (i.e. reducing the cloud to its filamentary structure). $AR_{\text{MA}}$ is then set as the ratio between the medial axis length, $L_{\text{MA}}$, and the medial axis width, $W_{\text{MA}}$, such that: $AR_{\text{MA}} = L_{\text{MA}} / W_{\text{MA}}$. $L_{\text{MA}}$ is simply the length of the determined medial axis, and $W_{\text{MA}}$ is twice the average distance from the medial axis to the cloud's external edge. The process of retrieving the medial axis fails when a cloud is too small (not enough pixels to erode away until only the skeleton remains); we set \textit{AR\_MA} to 1 for these cases. We do not attempt to retrieve filamentary structures for "fluffy", diffuse clouds - i.e. clouds that do not pass our robust background cut (further explained in following paragraphs) - in order to economise computational time; \textit{AR\_MA} is set 0 for these.

\subsection{Masses and surface densities}
\label{app:mass_sd_unc}

The total "flux" of a cloud (i.e. the sum of each pixel's gas mass surface density within a cloud, \textit{Sigma\_tot}) is computed by \texttt{ASTRODENDRO} using the bijection paradigm (see \citealt{rosolowsky_structural_2008}). The average surface density of each cloud, \textit{Sigma\_avg}, is then estimated by taking the total sum of surface densities within the cloud (\textit{Sigma\_tot}) and dividing it by the cloud's footprint area (\textit{Area\_exact}). Similarly, the peak surface density for each entry in the catalogue, \textit{Sigma\_peak}, is simply the highest surface density observed within a cloud. The mass of the cloud, \textit{Mass}, is then estimated as $M = \Sigma_{\text{avg}} \, A$ (i.e. average surface density of cloud multiplied by its area).  

In Paper I, we quantified the uncertainty of our opacity estimates through $10^4$ Monte Carlo realizations for each pixel in our gas surface density map. Our science sub-sample, which holds only clouds with average surface density above $10\,\sunpc$, has a maximum relative uncertainty of 45\%. Above $14\,\sunpc$ (the median cloud surface density across our molecular sub-sample, see Table \ref{tab:cloud_properties}), the maximum relative error drops below 30\%. It is also possible to obtain the relative uncertainty of masses and surface densities for each cloud in our catalogue. We compute the ratio between the total absolute error of the cloud (i.e. sum of the Monte Carlo mass/surface density uncertainties of each pixel inside the cloud in quadrature) and the total mass/surface density of the cloud. Each cloud's relative error on the mass and surface density is listed in the catalogue under \textit{Rel\_err}. In Paper I, we also determined the maximum surface density we are able to measure reliably given photometric noise, which has little impact in our cloud catalogue. In fact, out of the 13258 clouds that constitute our molecular sub-sample, only 27 MCs have more than 30\% of their area containing pixels where the surface density exceeds the maximum measurable surface density.

\subsection{Additional Tags}

\begin{figure}
    \centering
    \includegraphics[width=0.4\textwidth]{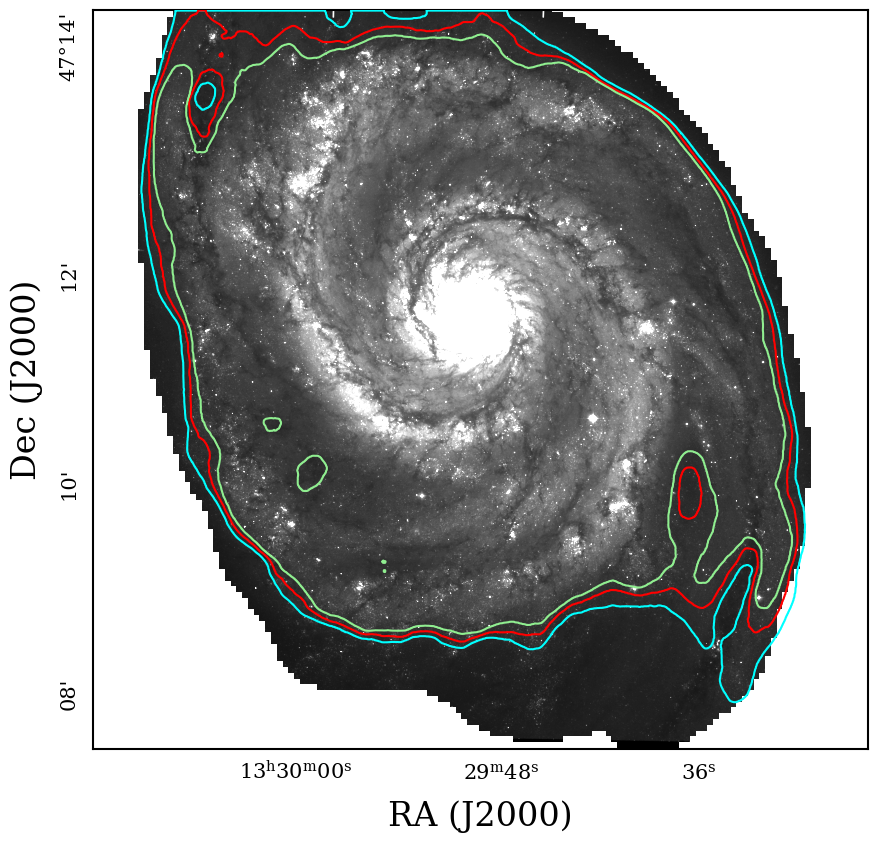}
    \caption{Original HST V-band image of M51. The overlaid contours correspond to 0.08, 0.09 and 0.1\,e$^-$/s levels (green, red, and blue respectively) in our stellar distribution map from Paper I.}
    \label{fig:I0_cut}
\end{figure}

In our analysis we only consider a subset of our full sample where we are more certain the clouds are real and dominantly molecular. As described in §\ref{sec:science_cut}, we consider only clouds that have a footprint area bigger than 3 resolution elements (flagged with \textit{Size\_cut}=1), are above the molecular surface density threshold, $\Sigma> 10\,\sunpc$ (\textit{Molecular\_cut}=1), and are against a robust stellar background (\textit{Robust\_bg}=1). The last flag is necessary because our technique retrieves extinction features through comparison with a modelled stellar distribution. Consequently, in regions where the stellar distribution is faint, the structures seen in extinction might not be real and are instead artefacts of our choice of background. The cloud shown in Fig.~\ref{fig:I0_cut_mc} (ID: 3) is an example of such a structure. Although its average surface density is above our molecular threshold ($\Sigma_{\text{avg}} \sim 10.9 \, \sunpc$) and its size is above 3 resolution elements ($A \sim 6.4\times10^{3}$ pc$^2$), it is not likely to be a real molecular cloud. In fact, almost 41\% of the pixels within this object have a measured surface density above the maximum surface density we can reliably measure (as explained in §\ref{app:mass_sd_unc}). This cloud borders the edge of M51 where there are not many stars that allow us to retrieve a reliable estimate of the stellar distribution ($I_0$), which is instrumental for our extinction technique (see Paper I for details). We therefore apply a robust $I_0$ cut to rule out these diffuse structures. Figure \ref{fig:I0_cut} shows the original HST V-band image with a choice of $I_0$ contours overlaid. Taking too large of a cut (e.g. $I_0 = 0.1$\,e$^{-}$/s, shown in green) rules out faint regions within the galaxy itself (which may be real), not just in the outskirts. Taking too little of a cut (e.g. $I_0 = 0.08$\,e$^{-}$/s, shown in blue) will not sufficiently exclude all faint locations. Our adopted $I_0$ threshold ($I_0 = 0.09$\,e$^{-}$/s, shown in red) seems like an adequate choice of cut where most of the galaxy is still considered and the regions without much stellar light are dismissed.

\section{Caveats in aspect ratio metrics}
\label{sec:appB}

\begin{figure}
    \centering
    \includegraphics[width=0.4\textwidth]{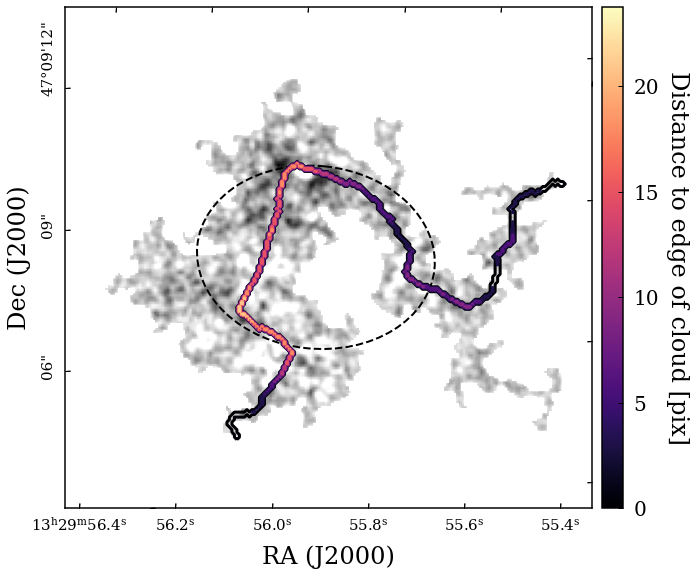}
    \caption{Example molecular cloud in our sample (ID: 4620) for which the aspect ratio from the medial axis and from the moments (see Appendix \ref{sec:appA}) differ significantly. The dashed black line ellipse represents the moments of the structure from which the moment aspect ratio, $AR_{\text{a/b}}$, is derived. The coloured line represents the medial axis of the cloud, where each pixel is colour-coded with the corresponding distance to the edge of the cloud. The background grey scale illustrates the surface densities computed from our extinction technique within the cloud's mask as defined by \texttt{SCIMES}. For this particular cloud, $AR_{\text{a/b}} = 1.3$ whilst $AR_{\text{MA}} = 15.1$.}
    \label{fig:medial_axis}
\end{figure}

It is useful to systematically study cloud morphologies, as a shape of the cloud may be linked or dictated by the dynamics of the surrounding medium. The simplest technique often employed is determining a cloud's aspect ratio. In a simplistic view, an aspect ratio would allow us to distinguish between "spherical" clouds and "filamentary" clouds. One way to estimate the aspect ratio of a cloud is through its moments where the structure is approximated by an intensity-weighted ellipse and the semi-major and semi-minor axis are then determined ($a$ and $b$, respectively), with the aspect ratio then being defined as $AR_{\text{a/b}} = a / b$. Another way to estimate the aspect ratio of a cloud is through the medial axis, $AR_{\text{MA}}$. This is a more geometrical approach by nature; it does not impose an elliptical structure and it is not weighted by intensity, instead it only takes the cloud's footprint mask into account to find the longest running spine which sits furthest away from the cloud edges. However, both of these metrics have issues when the morphology of a cloud becomes more complex, and also behave differently with different morphologies.

For example, for the cloud depicted in Fig.~\ref{fig:medial_axis}, with the moments approach we retrieve an $AR_{\text{a/b}}$ of 1.3, suggesting that we are dealing with a fairly circular cloud, even though it is clear from the figure that this is not the case. On the other hand, the medial axis aspect ratio $AR_{\text{MA}}$ has a value of 15.1, suggesting that this cloud is highly filamentary in nature. However, upon visual inspection, this MC is perhaps somewhere in between, and better classified as a ring-like cloud rather than a true filamentary structure. Thus while the aspect ratio can be used as a first glance at overall trends, any conclusions need to be carefully considered, as a more robust classification is needed in order to differentiate between real elongated structures and ring-like (or other complex morphologies) MCs.

%%%%%%%%%%%%%%%%%%%%%%%%%%%%%%%%%%%%%%%%%%%%%%%%%%

% Don't change these lines
\bsp	% typesetting comment
\label{lastpage}
\end{document}